# Impact Craters on Pluto and Charon Indicate a Deficit of Small Kuiper Belt Objects

K. N. Singer[1*], W. B. McKinnon[2], B. Gladman[3], S. Greenstreet[4], E. B. Bierhaus[5], S. A. Stern[1], A. H. Parker[1], S. J. Robbins[1], P. M. Schenk[6], W. M. Grundy[7], V. J. Bray[8], R. A. Beyer[9,10], R. P. Binzel[11], H. A. Weaver[12], L. A. Young[1], J. R. Spencer[1], J. J. Kavelaars[13], J. M. Moore[10], A. M. Zangari[1], C. B. Olkin[1], T. R. Lauer[14], C. M. Lisse[12], K. Ennico[10], New Horizons Geology, Geophysics and Imaging Science Theme Team[†], New Horizons Surface Composition Science Theme Team[†], New Horizons Ralph and LORRI Teams[†].

[1]Southwest Research Institute, Boulder, CO 80302, USA. [2]Department of Earth and Planetary Sciences and McDonnell Center for the Space Sciences, Washington University, St. Louis, MO 63130, USA. [3]University of British Columbia, Vancouver, BC V6T 1Z4, Canada. [4]Las Cumbres Observatory, Goleta, CA 93117, USA, and the University of California, Santa Barbara, CA, 93106 USA. [5]Lockheed Martin Space Systems Company, Denver, CO 80127, USA. [6]Lunar and Planetary Institute, Houston, TX 77058, USA. [7]Lowell Observatory, Flagstaff, AZ 86001, USA. [8]University of Arizona, Tucson, AZ 85721, USA. [9]Carl Sagan Center at the Search for Extraterrestrial Intelligence (SETI) Institute, Mountain View, CA 94043, USA. [10]National Aeronautics and Space Administration (NASA) Ames Research Center, Space Science Division, Moffett Field, CA 94035, USA. [11]Massachusetts Institute of Technology, Cambridge, MA 02139, USA. [12]Johns Hopkins University Applied Physics Laboratory, Laurel, MD 20723, USA. [13]National Research Council of Canada, Victoria, BC, Canada. [14]National Optical Astronomy Observatory, Tucson, AZ 26732, USA. *Corresponding author email: ksinger@boulder.swri.edu. [†]The full list of team members is available in the online version of the supplement.

**Abstract**: The flyby of Pluto and Charon by the New Horizons spacecraft provided high-resolution images of cratered surfaces embedded in the Kuiper belt, an extensive region of bodies orbiting beyond Neptune. Impact craters on Pluto and Charon were formed by collisions with other Kuiper belt objects (KBOs) with diameters from ~40 kilometers to ~300 meters, smaller than most KBOs observed directly by telescopes. We find a relative paucity of small craters $\lesssim$13 kilometers in diameter, which cannot be explained solely by geological resurfacing. This implies a deficit of small KBOs ($\lesssim$1 to 2 kilometers in diameter). Some surfaces on Pluto and Charon are likely $\gtrsim$4 billion years old, thus their crater records provide information on the size-frequency distribution of KBOs in the early Solar System.

**One Sentence Summary:** Pluto and Charon crater data reflect the small Kuiper belt object population, constraining models of planetary accretion and evolution.

The size-frequency distribution (SFD) of Kuiper belt objects (KBOs) with diameters (*d*) smaller than 100 km provides empirical constraints on the formation and evolution of bodies in the region beyond Neptune. The smaller bodies in the Kuiper belt are remnants of Solar System evolution and may have persisted largely unaltered for the past 4 billion years or longer (e.g., *7*). These objects also form an impactor population, producing craters seen on the younger surfaces of outer Solar System bodies (e.g., *5, 6, 8-10*). Constraining the primary impactor KBO size distribution can aid understanding of surface and interior evolution on these worlds. The KBO size distribution carries implications for whether comets (such as 67P/Churyumov–Gerasimenko) are primordial or substantially altered by collisions over their lifetimes (e.g., *11, 12, 13*), as well as for the number of small bodies discovered by telescopic campaigns (e.g., *17*).

Telescopic observations of smaller KBOs are difficult, so limited numbers <100 km in diameter have been confidently identified (e.g., *18, 19-21*). With few direct observations, predictions of the KBO population generally extrapolate from *d*~100 km objects down to smaller sizes by using a variety of plausible SFDs (e.g., *5*). Some predictions are informed by the impact crater SFDs observed on the icy satellites of Jupiter and Saturn, which reflect various combinations of impactor populations (*6, 14, 22*). Accretion and dynamical models of Kuiper belt formation predict either (i) many small and midsized objects ($d \lesssim 100$ km) form by hierarchical coagulation followed by runaway growth and collisional evolution (*3*), or (ii) far fewer small objects form when streaming or other aero-gravitational instabilities preferentially create larger bodies during the lifetime of the solar nebula (e.g., *23, 24-26*). The New Horizons flyby of the Pluto system (*27*) provides an opportunity to characterize the KBO population at small sizes by examining impact craters in the system.

**Impact crater distributions**

New Horizons imaged ~40% each of the surfaces of Pluto (~$7 \times 10^6$ km$^2$) and Charon (~$1.8 \times 10^6$ km$^2$), hereafter P&C, at moderate to high resolution during the 2015 flyby; these areas are referred to as their encounter hemispheres. The observed image scales range from 76–850 m px$^{-1}$ for Pluto, and from 154–865 m px$^{-1}$ for Charon, with the higher resolution datasets covering smaller areas (table S1; figs. S1 and S2). We mapped craters on each image sequence or scan of approximately consistent pixel scale independently, to allow for comparisons across image sets (*1*).

We consider craters in several broad regions on P&C, roughly delineated based on morphologic or albedo criteria (figs. S1 and S2**)**. The western and eastern sides of Pluto's encounter hemisphere are geologically different, and although terrains generally fall into broad latitudinal bands, they are not continuous across the encounter hemisphere. We divide Charon into two regions, the northern terrains (Oz Terra) and the southern Vulcan Planitia (VP; Fig 1), noting that some place names in this report are informal. VP is a large plain covering the southern portion of Charon's encounter hemisphere, which appears to have been completely (or nearly completely) resurfaced through icy volcanism, likely early in Charon's history (*28*). We focus on VP as it is highly favorable for crater mapping due to i) near-terminator lighting that emphasizes topography such as crater rims, ii) a mostly smooth initial surface, which aids in feature discrimination, and iii) a crater areal density far from saturation (i.e., craters with their ejecta are not strongly overlapping, which could obliterate and/or obscure one another). We prefer craters on VP as the clearest representation of the impactor flux, because these craters appear little modified or interrupted by younger craters or any subsequent geologic activity (Fig 1).

We quantify the crater populations and examine their geologic context to obtain information about the small body population

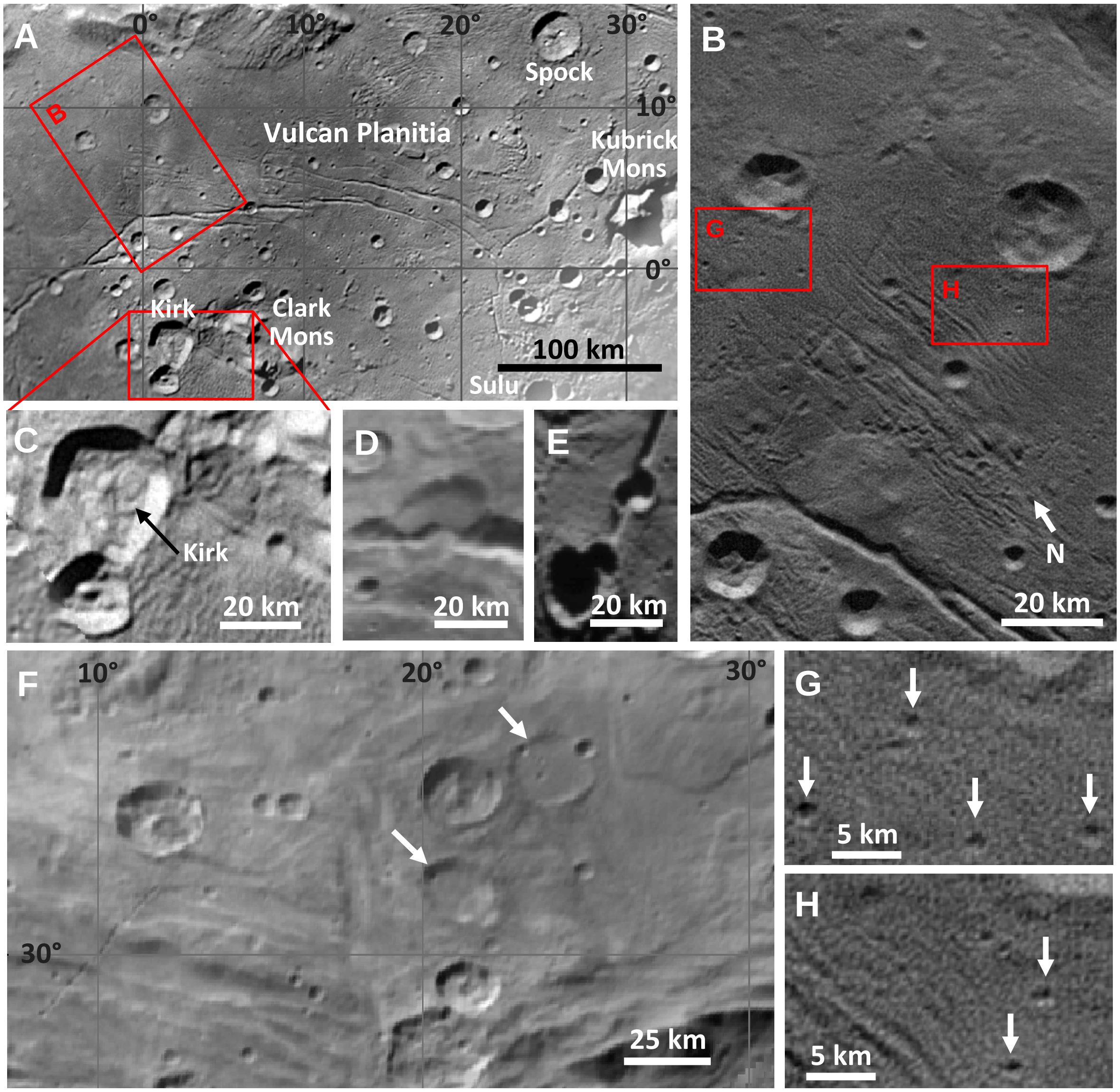

**Figure 1. Geology of Charon.** (**A**) Vulcan Planitia (VP); (**B**) high resolution view of VP with smoother regions; (**C**) craters with hummocky floors; (**D–E**) the only two craters on VP that may have been cut by faults, or alternatively their formation was affected by pre-existing faults; (**F**) partially filled craters (arrows) just north of the large chasmata separating the main expanse of VP from the northern terrains (Oz terra); and (**G–H**) views of small craters on VP. Arrows indicate small craters that were mapped, but only those larger than 1.4 km (10 pixels) were included in the analysis. Image resolutions and IDs (see Table S1): (A) mosaic of PELR_C_LORRI (865 m px$^{-1}$) and PELR_C_LEISA_HIRES (234 m px$^{-1}$); (G, D, H) PELR_C_MVIC_LORRI_CA (154 m px$^{-1}$); (C, D, E, F) PEMV_C_MVIC_LORRI_CA (622 m px$^{-1}$). Images are shown north upwards except B, where the orientation is indicated by the north arrow. Red rectangles indicate areas enlarged in other panels.

(KBOs) that impacted P&C. The differential number of craters per diameter ($D$) bin in a crater SFD is expressed as $dN/dD \propto D^q$ (*2, 29*), where $q$ is the power-law index or log-log slope. We use the symbol $D$ to refer to crater diameters, and the symbol $d$ to refer to impactor diameters. The relative or R-plot SFD divides this by $D^{-3}$ (*1*) such that a differential distribution with $q = -3$ is a horizontal line, whilst $q = -4$ and $-2$ form lines that slope downward and upward with increasing *D*, respectively (Fig. 2). The R-plot helps distinguish changes in slope as a function of diameter and between crater populations. We normalize the number of craters per bin by the mapped area to give the density of craters per size bin (*1*).

SFDs for both Pluto and Charon exhibit differential slopes ($q$) near −3 for $D \gtrsim 15$ km to $D$ ~100 km (there are few craters larger than this on either body) (Fig. 2; table S2). However, below ~15-km-diameter craters the SFDs for both P&C transition to a much shallower slope. Vulcan Planitia, our preferred crater-mapping surface (see above and discussion of geologic context below) has $q \sim -1.7$.

**Geological processes on Pluto and Charon**

Pluto has had a complex geologic history (*27*, *28*). The differing crater densities across Pluto and the range of SFDs in Fig 2 indicate that the geology influences crater modification and retention (as it does on Mars (*30*)). However, we cannot identify any geological resurfacing that has preferentially erased Pluto's small craters ($D \lesssim 15$ km) in all areas. Geologic processes that could substantially modify craters on Pluto are: [1] viscous relaxation of topography, [2] tectonics, [3] mass wasting, [4] glacial erosion by surface volatile ice ($N_2$, CO, $CH_4$) flow, [5] mantling by atmospheric fall-out or volatile ice deposition during Pluto's seasonal or climatic cycles, [6] sublimation erosion, [7] subsurface convective activity, or [8] icy volcanism as cryovolcanism (supplementary text).

Viscous relaxation [1] preferentially erases larger craters except in unusual circumstances (*29*, *31*). Viscous relaxation, tectonics, mass wasting, and glacial activity [1–4] rarely completely erase craters and there is no geomorphic evidence for these processes modifying craters on Pluto, which would produce a range of degradation states. There is evidence that processes [5–8] have modified craters (or at least terrains) on Pluto, and could potentially erase more small craters than large ones. Of these four processes, there is the most evidence for mantling, or covering of a terrain with a new layer of material. This is most clearly observed in Pluto's northern terrains (Lowell Regio, fig. S3A), where partially filled larger craters and general softening of topography imply that smaller craters may be completely buried (*28*). Some smooth terrains are present in Pluto's dark equatorial band (Cthulhu, fig. S3B) and may indicate localized in-filling. However, the topography there is generally not softened to the same degree as in Lowell, and Cthulhu does not show as many smooth-floored, nearly-filled craters as seen in the north. On Pluto, cryovolcanism has possibly occurred in specific regions (e.g., Wright Mons), convection in Sputnik Planitia's nitrogen-ice-rich plains, and sublimation erosion in other regions (e.g., Piri Planitia and the "bladed" terrain); these appear to have removed or buried all craters in their respective regions, large and small (*28, 32*) (fig. S3D-G).

On Charon's Vulcan Planitia, the main resurfacing process that has affected large surface areas is cryovolcanism. Charon has no detectable atmosphere (*33*), and there is no evidence for past glacial activity involving volatile ices (or even their presence on the surface) as there is on Pluto. Volcanic and tectonic activity has occurred since the emplacement of the main VP unit, but there is little evidence that this later activity has embayed or directly modified individual craters (Fig. 1). That is, none of the craters on the main VP area are partially filled, embayed, or breached with smooth, flat floors (in the manner seen on the lunar mare (*34*)). A few partially filled craters occur north of the large tectonic scarps that separate the main area of VP from Oz Terra (Fig. 1F); this embayment may have been contemporaneous with the emplacement of the VP to the south or could represent a later episode. Some large craters on VP have hummocky floors (Fig. 1C). At the available image resolutions, these hummocks could be viscous flow features, or they may be landslide material, but they appear similar to mass movement (e.g., landslides and debris flows) and collapse features seen in crater floors on other worlds (e.g., the Moon; fig. S4 (*34*)). A few smooth patches (~50-100 km across) with even fewer craters are found in Vulcan Planitia itself, suggesting possible resurfacing by flows after the main resurfacing event(s) (Fig. 1A; fig. S5). However, these smooth areas are generally devoid of large craters (with only a few smaller craters throughout), and thus do not indicate that small craters were preferentially erased in these areas.

Overall, the break in logarithmic slope and deficit of smaller craters on P&C appears to be a primary characteristic of the impacting body population because: i) no process or combination of processes appear to have preferentially erased small craters across all areas of P&C, and ii) the SFD break in slope or turnover occurs at approximately the same diameter ($D$ ~10–15 km) on both P&C and across varied terrains or subsets of terrains (fig. S6). The same or similar shallow slope is seen for equivalent (scaled) crater sizes on young terrains on Jupiter's moon Europa, and on (relatively) younger impact basins on

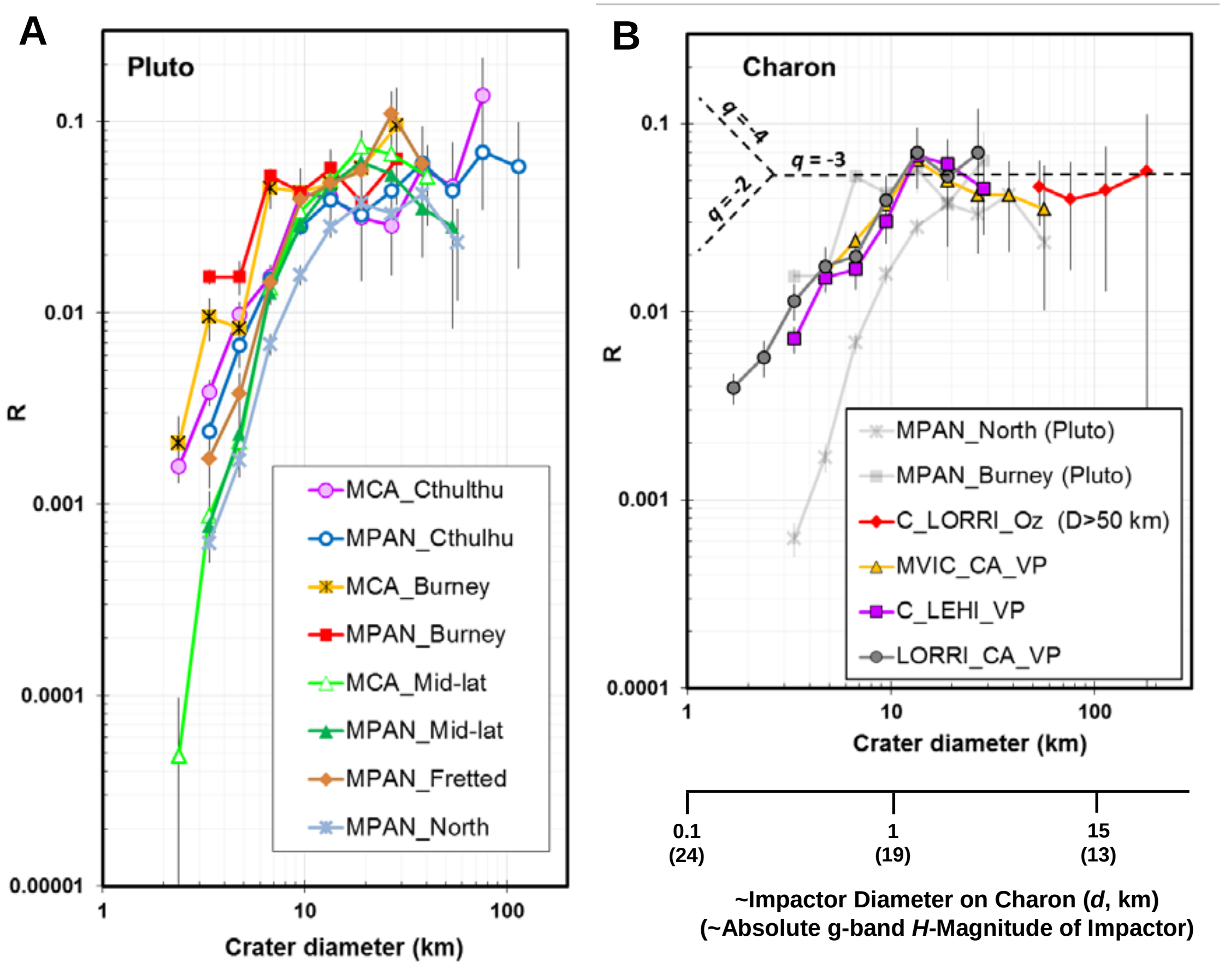

**Figure 2. Regional Crater Size-frequency Distributions.** (**A**) Pluto and (**B**) Charon SFDs shown as relative differential, or R-plots, which display crater spatial densities normalized to a $D^{-3}$ distribution (see text), for areas with more than 10 craters. In panel B, the highest and lowest Pluto SFDs are overlain in light grey for comparison with the Charon SFDs. Legend titles refer to the observation name and mapped region, as listed in table S1 and shown in figs. S1 & S2. For reference, the additional axis below panel B shows an approximate impactor diameter (based on water ice density for impactor and target) and approximate equivalent g-band H-magnitude, a measure of brightness used in telescopic observations of KBOs (*1*). Dashed lines indicate slopes of differential SFD power law $q$ values of -2, -3 and -4. Error bars represent Poisson statistical uncertainties (*2*).

Ganymede and Callisto (*6*) (fig. S7).

With regard to the last point, the icy satellites of Jupiter and Saturn are thought to be cratered by objects ultimately derived from the Kuiper belt (*6*). Although young surfaces on Europa and Ganymede were by definition resurfaced in the past, the existing craters on these younger surfaces show few signs of geological disruption, and craters above ~2-km in diameter are thought to represent the primary impact flux at Jupiter and Saturn (*6, 14, 35, 36*). The shallow SFD slope found at small diameters on some of these icy satellites also impies a dearth of small comets or other KBOs) (*6*), and is thus consistent with our results.

The surfaces of P&C do not exhibit any obvious secondary craters, which are smaller craters formed by ejected fragments from a primary impact. Any dispersed (as opposed to clustered and thus obvious) secondary craters, or craters from collisionally derived debris within the Pluto system, would tend to increase the number of small craters relative to large ones and lead to a steeper crater SFD (*37-39*). These populations thus do not appear to play a role at the scale of our measurements, because we see no morphological or spatial patterns that would indicate they are present and we observe shallow SFD slopes. If impactors internal to the P&C system do contribute, this would require the primary, heliocentric, small KBO impactor population SFD slope to be even shallower to match the observed crater SFDs.

## Implications for Kuiper belt populations

The craters on Charon's Vulcan Planitia have a break in the slope (or turnover) close to 13 km (fig. S8). Similar breaks are seen in regions on Pluto (Fig 2), but we adopt the VP turnover diameter for the remainder of the analysis because we consider the craters on VP the best representation of the primary impactor population (as discussed above). A crater diameter of ~13 km on Charon corresponds to an impactor diameter (*d*) of ~1–2 km (*1*). The crater data from VP indicate a shallow slope of $q \sim -1.7 \pm 0.3$ for craters from ~1–13 km in diameter (using the mean and median value of the weighted slope fits in table S2 and the largest error estimate), corresponding to impactors from ~100 m to 1 km in diameter. We will refer to this break in slope at $d \sim 1$–2 km KBOs as the "elbow" as there is already a slope break known for $d \sim 100$ km KBOs (*18, 40-44*) dubbed the "knee". Note there are too few large (>100-km diameter) craters on P&C to observe the knee slope change itself in the crater population, though the dearth of such craters is consistent with the steepening of the SFD at large *d*.

The elbow is located at an absolute g-band magnitude $H_g \sim 18$, and the knee at $H_g \sim 9$, assuming a geometric g-band albedo of 0.05 to convert between size and magnitude estimates(*1*). The crater SFD slopes (*q*) translate to $H_g$ differential magnitude distribution slopes ($\alpha$) via $q = -(5\alpha + 1)$ (*5*). The crater SFD slopes thus equate to $\alpha$ of ~0.4 ± 0.04 and ~0.15 ± 0.04, respectively, for (uniform albedo) impactors larger and smaller than the elbow. The Outer Solar Systems Origins Survey has reported $\alpha = 0.4$ ($q \sim -3$) for KBOs $d < 100$ km (*45*). However, Earth-based observational searches have not yet probed sufficient numbers of small KBOs to test the additional elbow break at $d \sim 1$–2 km (or $H \sim 18$).

The craters on P&C represent impacts from a mix of different Kuiper Belt dynamical subpopulations, with four contributing the majority of impactors: the stirred component of dynamically cold classical KBOs, the dynamically hot classical KBOs, other 3:2 Neptune-resonant KBOs (also known as plutinos), and the outer classical KBOs (including objects whose orbits are detached from Neptune interaction) (*4, 5, 14*). Figure 3 compares the SFDs from Pluto and Charon with predictions constructed from several impactor flux models for the outer solar system (*4-6, 14*) and one model based on a collisional evolution of the asteroid belt (*15, 16*). The G16 knee model (*4*) uses Earth-based telescopic constraints to set the SFD slope and number for Kuiper belt objects larger than ~100 km in diameter, and bends to a shallower slope (for all subpopulations) below that size (*5*, *18*, *19*). Although the absolute impactor flux levels carry substantial uncertainty (*4*, *5*), the three different age predictions shown in Fig. 3A for the knee model suggest that the surface of Charon is ancient, close to 4 Gyr (*28*).

In Fig. 3B, the Z03 model is based on crater distributions on the icy satellites of Jupiter. The Z03 4-Gyr prediction (*6*) follows the overall shape of the Charon crater SFDs and is similar in overall crater density to the knee model between 20 km ≲ *D* ≲ 100 km (*6*). The BD15 model (*14*) predicts crater densities about an order of magnitude below the knee model (for 4 Gyr), but the −2 slope motivated by the icy jovian satellite craters is again more similar to that seen for P&C's smaller craters. The P&C crater data are inconsistent with the S13 model (Fig. 3A), which is calibrated using two stellar occultations by small (*d* ~ 0.5 km) KBOs (*46, 47*), after conversion to equivalent Charon crater densities (*4*). The slope from the S13 collisional evolution model is steep and does not consistently match the P&C crater densities over any diameter range (for surface ages of 100 Myr, 1 Gyr or 4 Gyr, see Fig. 3A).

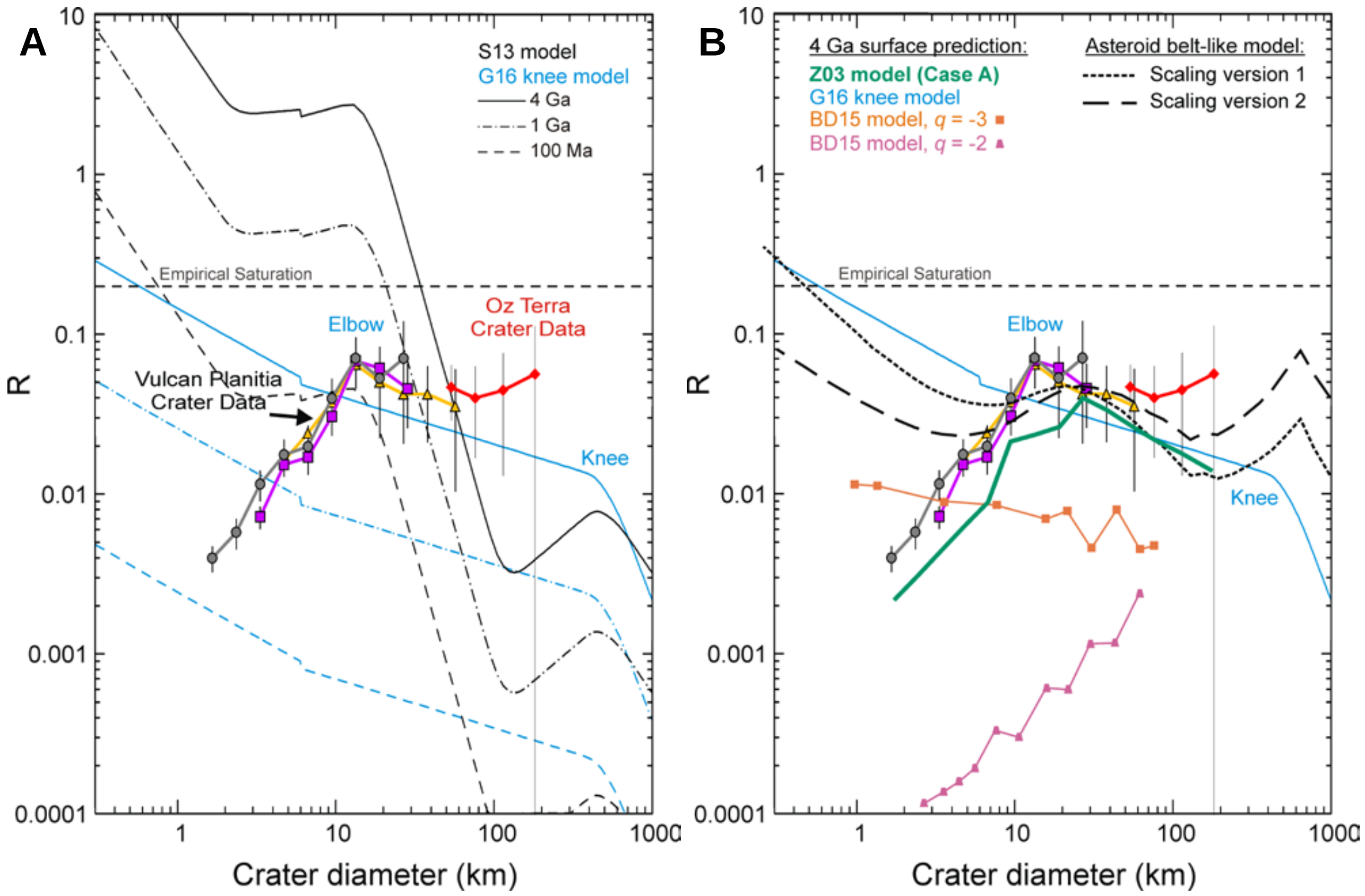

**Figure 3. Charon crater SFDs compared to impactor population models.** (**A**) Charon crater data (points, same as in Fig. 2) along with predictions for crater populations from two impactor flux models S13 (*3*) and G16 (*4*, *5*) (curves), shown for three different age surfaces. The empirical crater saturation density as calculated in G16 (*4*, *5*) is also shown for reference, where typically the density of craters does not increase because subsequent impacts erase previous craters while forming new ones. (**B**) The same crater data shown with three more models of impactor flux, Z03 (*6*), G16 (*4*, *5*) and BD15 (*14*), all for a 4 Gyr old surface . The knee model from panel A is included for reference, as are two different scalings of a collisional evolution model of the asteroid belt (*1*, *15*, *16*). The crater data do not match the slopes predicted from collisional evolution models (in which collisions between KBOs would produce copious fragments) and does not resemble the asteroid belt at small sizes. The Z03 model, based on craters on the young surfaces of Europa and Ganymede (*6*), is the closest match to the Charon craters (*D* < 10 km) over the entire data range.

The KBO SFD derived from P&C is not consistent with a population in traditional collisional equilibrium where destructive collisions would produce an average SFD slope of *q* ~ −3.5 below some scale, with undulations or waves superimposed (*1*). Collisional models can match the asteroid belt SFD (*48*) and have been applied to explaining the Kuiper belt knee (*49*). It is possible the KBO population may have reached a collisional equilibrium state over a limited size range (the slopes for craters between ~10- and 100-km diameter are consistent with such collisional models), but with smaller impactors (*d* < 1 km) lost by some other process(es). If small objects have a low internal strength, some collision models can produce shallow slopes by destroying objects smaller than 1 km (*50*). Stirring by Neptune can also affect the predicted SFD structure (*51*). Thus it is possible that different KBO structural properties or dynamical influences may result in collisional outcomes different from those in the asteroid belt. Nevertheless, the differential *q* ~ −1.7 slope over at least one order of magnitude in crater size is difficult to reproduce in models.

Alternatively, the SFD shape we see on P&C may represent a more primordial (less collisionally evolved) KBO impactor population, possibly a remnant accretional size distribution (*1*), which would constrain early Solar System processes. If the shallower slope for small KBOs is primordial, it may be more consistent with dynamical models of gravitational instabilities causing rapid growth to larger objects (*11, 23-25*). These models produce fewer small bodies, which could also result in fewer collisions overall, leading to a lower production rate of collisional fragments. However, collisions of even a few large objects would produce a collisional tail of small bodies and cause the overall SFD to evolve towards an equilibrium slope (*q* ~ −3–4) (e.g., *48*). Models of streaming instabilities (*52*) followed by gravitational collapse produce size distributions with *q* ~ −2.8 down to *D* ~ 80 km (the smallest objects numerically resolved) (*25, 26, 53*). Gravitational collapse should become less efficient for smaller-mass clumped regions at some characteristic scale or scales, which could potentially explain the shallow SFD slopes (*q* values) we observe (*54*) though higher-resolution studies are likely needed to fully address this issue. There may be more than one combination of processes that can produce the observed KBO SFD. If there are fewer small objects in the outer solar system, the probability of collisions is lower, and comets such as 67P/Churyumov–Gerasimenko are less likely to be affected by many catastrophic collisions throughout their lifetimes (*11, 12*) (supplementary text). By providing constraints on a size range of KBOs not currently accessible by telescopes, the New Horizons crater data can help discriminate between models of accretion, evolution, and/or emplacement of the Kuiper belt.

## References and Notes

1. A. Morbidelli et al., in The Solar System Beyond Neptune, M. A. Barucci, Ed. (2008), pp. 275–292.
2. K. Zahnle, P. Schenk, H. Levison, L. Dones, Icarus 163, 263–289 (2003).
3. P. M. Schenk, C. R. Chapman, K. Zahnle, J. M. Moore, in Jupiter, F. Bagenal, T. E. Dowling, W. B. McKinnon, Eds. (Cambridge Univ. Press, New York, 2004), pp. 427–456.
4. L. Dones et al., in Saturn from Cassini-Huygens, M. K. Dougherty, Ed. (2009), pp. 613–635.
5. M. Kirchoff et al., in Enceladus and the Icy Moons of Saturn, P. Schenk, C. Howett, Eds. (University of Arizona Press, 2018), pp. 267–306.
6. S. Greenstreet, B. Gladman, W. B. McKinnon, Icarus 258, 267–288 (2015).
7. B. J. R. Davidsson et al., Astron. Astrophys. 592, A63 (2016).
8. M. Jutzi, W. Benz, A. Toliou, A. Morbidelli, R. Brasser, Astron. Astrophys. 597, A61 (2017).
9. A. Morbidelli, H. Rickman, Astron. Astrophys. 583, A43 (2015).
10. M. J. Lehner et al., in SPIE Astronomical Telescopes + Instrumentation (SPIE, 2016), vol. 9906, pp. 10.
11. G. M. Bernstein et al., Astron. J. 128, 1364–1390 (2004).

12. W. C. Fraser, M. E. Brown, A. Morbidelli, A. Parker, K. Batygin, Astrophys. J. 782, 100 (2014).
13. M. Alexandersen et al., Astron. J. 152, 111 (2016).
14. M. W. Buie et al., Paper presented at the 49th Annual American Astronomical Society Division for Planetary Sciences Meeting, Provo, UT, 1 October 2017.
15. E. B. Bierhaus, L. Dones, Icarus 246, 165–182 (2015).
16. D. A. Minton, J. E. Richardson, P. Thomas, M. Kirchoff, M. E. Schwamb, paper presented at the Asteroids, Comets, Meteors 2012 Conference, Niigata, Japan, 1 May 2012, no. 6348.
17. H. E. Schlichting, C. I. Fuentes, D. E. Trilling, Astron. J. 146, 36 (2013).
18. E. Chiang, A. N. Youdin, Annu. Rev. Earth Planet. Sci. 38, 493–522 (2010).
19. D. Nesvorný, A. N. Youdin, D. C. Richardson, Astron. J. 140, 785–793 (2010).
20. A. Johansen, M.-M. Mac Low, P. Lacerda, M. Bizzarro, Sci. Adv. 1, e1500109 (2015).
21. J. B. Simon, P. J. Armitage, R. Li, A. N. Youdin, Astrophys. J. 822, 55 (2016).
22. S. A. Stern et al., Science 350, aad1815 (2015).
23. Materials and methods are available in the supplementary materials.
24. J. M. Moore et al., Science 351, 1284–1293 (2016).
25. H. J. Melosh, Impact cratering: A Geologic Perspective (Oxford Univ. Press, New York, 1989), pp. 253.
26. Crater Analysis Techniques Working Group et al, Icarus 37, 467–474 (1979).
27. M. Gurnis, Icarus 48, 62–75 (1981).
28. S. A. Stern, S. Porter, A. Zangari, Icarus 250, 287–293 (2015).
29. W. B. McKinnon et al., Nature 534, 82–85 (2016).
30. S. A. Stern et al., Icarus 287, 124–130 (2017).
31. D. E. Wilhelms, The Geologic History of the Moon (1987), QB592.W5.
32. E. B. Bierhaus, C. R. Chapman, W. J. Merline, Nature 437, 1125–1127 (2005).
33. E. B. Bierhaus, K. Zahnle, C. R. Chapman, in Europa, R. T. Pappalardo, W. B. McKinnon, K. K. Khurana, Eds. (Univ. of Arizona Press, Tucson, 2009), pp. 161–180.
34. A. S. McEwen, E. B. Bierhaus, Annu. Rev. Earth Planet. Sci. 34, 535–567 (2006).
35. K. N. Singer, W. B. McKinnon, L. T. Nowicki, Icarus 226, 865–884 (2013).
36. R. A. Smullen, K. M. Kratter, Mon. Not. R. Astron. Soc. 466, 4480–4491 (2017).
37. W. C. Fraser, J. J. Kavelaars, Icarus 198, 452–458 (2008).
38. C. I. Fuentes, M. J. Holman, Astron. J. 136, 83–97 (2008).
39. B. Gladman et al., Astron. J. 122, 1051–1066 (2001).
40. A. H. Parker et al., paper presented at the 46th Lunar and Planetary Science Conference, Houston, TX, 1 March 2015, no. 2614.
41. C. Shankman, B. J. Gladman, N. Kaib, J. J. Kavelaars, J. M. Petit, Astrophys. J. Lett. 764, (2013).
42. C. Shankman et al., Astron. J. 151, 31 (2016).
43. S. Greenstreet, B. Gladman, W. B. McKinnon, Icarus 274, 366–367 (2016).
44. W. F. Bottke Jr et al., Icarus 175, 111–140 (2005).
45. W. F. Bottke Jr et al., Icarus 179, 63–94 (2005).
46. H. E. Schlichting et al., Astrophys. J. 761, 150 (2012).
47. H. E. Schlichting et al., Nature 462, 895–897 (2009).
48. W. F. Bottke et al., in Asteroids IV, P. Michel, Ed. (Univ. of Arizona Press, Tucson, 2015), pp. 701–724.
49. M. Pan, R. Sari, Icarus 173, 342–348 (2005).
50. S. J. Kenyon, B. C. Bromley, Astron. J. 143, 63 (2012).
51. S. J. Kenyon, B. C. Bromley, Astron. J. 128, 1916–1926 (2004).
52. A. Youdin, A. Johansen, Astrophys. J. 662, 613–626 (2007).
53. J. B. Simon, P. J. Armitage, A. N. Youdin, R. Li, Astrophys. J. 847, L12 (2017).
54. C. P. Abod, J. B. Simon, R. Li, P. J. Armitage, A. N. Youdin, K. A. Kretke, The mass and size distribution of planetesimals formed by the streaming instability. II. The effect of the radial gas pressure gradient. arXiv:1810.10018 [astro-ph.EP] (23 October 2018).

**Acknowledgments:** We thank the entire New Horizons team for their hard work leading to a successful Pluto system flyby. We thank three anonymous reviewers for helpful comments that improved this manuscript. We also thank M. Kirchhoff, K. Zahnle, B. Bottke, S. Marchi, S. Jacobson, D. Nesvorný, A. Schreiber, A. Youdin, and J. Simon for sharing their data and/or engaging in discussions. Funding: New Horizons team members gratefully acknowledge funding from the NASA New Horizons Project. B.G. acknowledges support of the Natural Sciences and Engineering Research Council of Canada. S.G. acknowledges support from NASA NEOO grant NNX14AM98G to Las Cumbres Observatory. Author contributions: K.N.S conducted the crater size-frequency analysis, coordinated the research, and co-wrote the paper. W.B.M., B.G., S.G., and E.B.B. contributed ideas and co-wrote the paper. S.A.S., A.H.P, S.J.R., P.M.S., W.M.G., V.J.B., R.A.B., R.P.B., H.A.W., L.A.Y., J.R.S., J.J.K., J.M.M, A.M.Z., C.B.O., T.R.L, C.M.L., and K.E. are New Horizons team members who contributed to the development of the paper through discussion and text suggestions, as well as contributing to the success of the New Horizons encounter with Pluto that enabled the data presented in this paper. Competing interests: There are no competing interests to declare. Data and materials availability: The data points in Figs. 2 and 3 are provided in Data S3. The data needed to derive the points in Figs. 2 and 3 are provided in Data S1 and S2. The specific image data used is described in Table S1. The LORRI data are archived in the Planetary Data System (PDS) Small Bodies Node at https://pds-smallbodies.astro.umd.edu/holdings/nh-p-lorri-3-pluto-v2.0/. MVIC data are available via the PDS at https://pds-smallbodies.astro.umd.edu/holdings/nh-p-mvic-3-pluto-v2.0/.

# Supplementary Materials for

## Impact Craters on Pluto and Charon Indicate a Deficit of Small Kuiper Belt Objects

K. N. Singer*, W. B. McKinnon, B. Gladman, S. Greenstreet, E. B. Bierhaus, S. A. Stern, A. H. Parker, S. J. Robbins, P. M. Schenk, W. M. Grundy, V. J. Bray, R. A. Beyer, R. P. Binzel, H. A. Weaver, L. A. Young, J. R. Spencer, J. J. Kavelaars, J. M. Moore, A. M. Zangari, C. B. Olkin, T. R. Lauer, C. M. Lisse, K. Ennico, New Horizons Geology, Geophysics and Imaging Science Theme Team, New Horizons Surface Composition Science

Correspondence to: ksinger@boulder.swri.edu

**This PDF file includes:**

Materials and Methods
Supplementary Text
Figs. S1 to S8
Tables S1 to S3
Captions for Data Files S1 to S3
References

**Other Supplementary Materials for this manuscript include the following:**
(available at www.sciencemag.org/content/363/6430/955/suppl/DC1)

Data Files S1 to S3 (csv files)

## Materials and Methods

Mapping methods and data display.

We present crater mapping results from both the LORRI (LOng Range Reconnaissance Imager (*55*)) framing camera, and the Ralph/MVIC (Multispectral Visual Imaging Camera (*56, 57*)) scanning imager. The images were converted to normalized reflectance values following the Planetary Data System (PDS) documented calibration for each instrument (*58*), a lunar-Lambert photometric correction was applied, and the images were mosaicked through the United States Geologic Survey (USGS) Integrated Software for Imagers and Spectrometers (*59*). Image pixels scales and areal coverage are given in table S1 and figs. S1–S2. Each image sequence or scan at a given scale was mapped and plotted independently for the regions covered by those images. Mapping was conducted on simple cylindrical projections for the regions below ~60° latitude and on polar stereographic projections for regions above ~60° latitude. The projection warp matches that used for the global mosaics publically released in July 2017 and available from the NASA PDS IMG Annex (*58*). The request IDs given for each dataset (table S1) are a unique identifier in the Planetary Data System image headers.

Mapping involves interpretations; we include features deemed to more likely be craters than any other geologic feature type. Fresh, obvious craters are relatively circular depressions with raised rims, signs of ejecta (although these are generally not obvious on Pluto), and/or classic interior morphologies as observed on other icy and rocky surfaces in the solar system (central peaks, wall terraces, etc.) (*29*). Stereo topography (*60, 61*) was reviewed for some ambiguous features. Crater mapping confidence levels were assigned to each feature as it was mapped as described in (*62*). Our crater measurements include all features whose consensus confidence level was greater than 50% of being an actual impact feature. Fig. S5 shows the craters mapped in the highest resolution image mosaic over Charon's Vulcan Planitia. Including all depressions that could have the slightest chance of being a crater (such as features that are depressions but are not very circular and lack raised rims) would raise the R-values in the smaller diameter bins by 20-30% at most, but would not affect the conclusions in this paper.

For all relative- or R-plots in this paper, each region is normalized to its own area and bin diamater widths are factors of √2, except for in some cases the largest diameter bin was increased in size to prevent bins with < 1 crater. Binned R-values are calculated from $N/(D_{mid}^{-3}\,[D_b - D_a]\,A)$ where $N$ is the number of craters per bin, $D_a$ and $D_b$ are the lower and upper bounds of the bin, $D_{mid}$ is the geometric mean of $D_a$ and $D_b$, and $A$ is the mapped unit's surface area. Many solar system crater differential distributions ($dN/dD$) display a log-log slope close to −3, thus the normalization by $D^{-3}$ highlights variation from this typical case. Poisson statistical uncertainties are assumed (*2*). Below we also use a smooth kernel density estimate to display the R-plot as an alternative form of binning the data, and to examine the location of the "elbow" slope break (fig. S8).

With the exception of Oz Terra on Charon (C_LORRI_Oz in Fig. 2B), the distributions are truncated below the diameter of ~7–10 pixels for each dataset, and thus the data shown are well above the resolution limit for crater detection. The lighting geometry in Oz Terra (small solar incidence angles, meaning no shadows) precludes easy identification of smaller craters, so we have selected a conservative $D > 50$ km subset for this region. The smallest craters presented here are from the highest resolution image strip (154 m px$^{-1}$) on Charon (Fig. 1B,G,H, figs. S5, S6), where a 1 km crater is ~6.5 pixels across. We truncate the R-plots below 10 pixels (for an ~1.4-km-diameter crater, which is the lower bound on a √2 bin) for this highest resolution strip (LORRI_CA_VP in Fig. 2) as a conservative limit, although smaller features can be seen. All three different resolution image sets covering subsets of VP show similar crater densities (R values) and similar overall SFD slopes for the smallest craters observed. We also examined some subsets of VP to test for specific regional differences in the slope. Although the eastern portion of VP has several deep depressions and much more topographic variation compared with the western portion, both the east and west portions of VP have similar SFDs (fig. S5). Additionally, in the highest resolution image strip we looked at the differences between smoother terrains and textured/pitted terrains (see fig. S5). The uncertainties in our crater diameter measurements are 10–15% or less, set by the 7–10 pixel scale of our smallest measured craters, decreasing for larger craters. See (*63*) for a general discussion. Uncertainties in measured areas (in ArcGIS) are negligible in comparison.

Crater to impactor size scaling.

We follow Schmidt-Housen-Holsapple impactor scaling (*64*, *65*). We use cold water ice parameters (representing a non-porous material) in the gravity regime, a transient crater depth-to-diameter ratio of 0.2, an impact angle of 45°, and the formulation for the final-to-transient complex crater conversion derived from crater shapes on icy satellites (*66*). This method is described in detail in previous papers (*38, 67*). We use an average impactor velocity ($U$) of 2 km $s^{-1}$ for Charon and 2.2 km $s^{-1}$ for Pluto (both equivalent to a velocity at infinity of 1.9 km $s^{-1}$ for the system). The parameters used for water ice are target and impactor density ($\rho$ and $\delta$, respectively) of 920 kg $m^{-3}$, scaling parameters $K_1$ = 0.095 and $K_2$ = 0.351, scaling exponents $\mu$ = 0.55 and $\nu$ = 0.33, and surface gravity for each world ($g$). In the gravity regime the scaling parameters are similar for all non-porous materials (e.g., rock versus ice), wherein material strength does not affect formation of the transient crater cavity.

Using the inputs above in the general equations for scaling, the impactor diameter ($d$) as a function of final crater diameter ($D$) for Pluto is:

$$d = 0.08D^{1.151} \quad \text{(S1)}$$

and for Charon:

$$\mathrm{d} = 0.07\mathrm{D}^{1.151} \quad \text{(S2)}$$

where $D$ and $d$ are in km. For reference, assuming a porous target material with a lower density (644 kg $m^{-3}$, for ice with 30% porosity), $\delta$ = 920 kg $m^{-3}$, scaling parameters $K_1$ = 0.132 and $K_2$ = 0, and scaling exponents $\mu$ = 0.41 and $\nu$ = 0.33, (the suggested parameters for a sand- or regolith=like material; e.g., (*64*)) yields a scaling law for Pluto of:

$$d = 0.14D^{1.088} \quad \text{(S3)}$$

and for Charon:

$$d = 0.13D^{1.088} \quad \text{(S4)}$$

As an example, a 10-km-diameter crater on Charon in a non-porous material is predicted to be formed by a ~1.3 km impactor, compared with a $D$ = 10 km crater formed in a porous material, which requires a 1.8 km impactor. Impactor sizes scale with assumed impactor and target densities as $(\rho/\delta)^{1/3}$ (*64, 65*). The scaling factor (SF) between crater diameters on other worlds ($D_{other}$) and the diameter of a crater the same size impactor would make on Charon ($D_{Charon}$) follows (for non-porous targets):

$$SF = \left(\frac{g_{other}}{g_{Charon}} * \frac{U_{Charon}^2}{U_{Other}^2}\right)^{0.216}$$

(S5)

$$D_{Charon} = SF * D_{other} \quad \text{(S6)}$$

where $D$ is in kilometers. Average impact velocities and scale factors for each world are given in table S3.

For the Z03 Case A in Fig. 3, we plot the predicted number of impactors per year multiplied by 4 Gyr (with no decay factor over the 4 Gyr) by differentiating equation 14 in (*6*) and dividing by $D^{-3}$, which produces a function whose slopes are piecewise continuous. In this case, conversion from impactor diameters to crater diameters was carried out using equations 5 and 6 in (*6*) and the parameters stated in that paper, so as to maintain the original predictions. We sampled the resulting SFD in R-values at the same crater diameter ($D$) bin spacing as the C_LORRI_CA and C_LORRI_Oz_Terra distributions for Charon.

<u>KBO/impactor diameter to absolute magnitude conversion.</u>

Absolute magnitude in g-band ($H_g$) as defined by the Sloan Digital Sky Survey has a central wavelength of 4686 Å. We convert from KBO diameter to $H_g$ using (*5*):

$$d \cong 100km\sqrt{\frac{0.05}{p}10^{0.2(9.16-H_g)}} \quad \text{(S7)}$$

where $p$ is the g-band albedo and a constant $p$ = 0.05 is used throughout this work. Converting size distribution slopes ($q$) to magnitude distribution slopes ($\alpha$) follows $q = -(5\alpha + 1)$, for the definition of $q$ used here (*5*). Higher albedos would imply smaller KBOs (keeping $pd^2$constant). If albedos are scale dependent (*68*), this can affect the calculation of the magnitude distribution slope ($\alpha$) but would not affect the crater scaling or calculation of the crater SFD slope.

<u>Asteroid belt comparison.</u>

Collisional evolution models of the asteroid belt (*15, 16*) can mimic the observed main asteroid belt populations. Models of growth and collisional evolution of small solar system bodies produce bumps/waves or changes in slope in the population SFDs due to various physical processes (*3, 15, 16, 48*). However, the models generally produce a fairly consistent slope across the population that evolves as collisions proceed, with the waves superimposed upon it (*48*), and that slope reaches a steady state or collisional equilibrium around $q \sim -3.5$. The wavy shape of the asteroid belt model includes two main "bumps". The model bump at larger sizes ($d$ ~ 100 km in the asteroid belt, which would create a crater on Charon of $D$ ~ 500 km), matches the observed large asteroid population and is thought to be a remnant of accretion and the primordial object SFD in the main belt (*15*), while the model bump at smaller sizes ($d$ ~ 3–4 km or a $D$~30 km crater on Charon) is a wave propagated by the strength-to-gravity transition for catastrophic asteroid breakup (at $d$ ~ 200 m for asteroids (*69*)).

We apply two different scalings to the asteroid belt model in order to compare to the P&C crater data. The first scaling ("version 1" in fig. S7B) takes the model size-frequency distribution and uses eq. 2 to determine crater sizes on Charon. This introduces a small change in the distribution slopes. The second version ("version 2" in fig. S7B) shifts the asteroids to the same sizes as version 1 but preserves the original model slopes. Both are arbitrarily shifted in R-value to yield a similar crater density as on Charon's encounter hemisphere.

The shape of asteroid-belt-like populations is different from that observed on Pluto and Charon. The clearest difference is the steep increase at small diameters (for the asteroid belt, interpreted as a product of collisional evolution), but missing from the P&C data. It is possible that we are not able to observe small enough craters on P&C to detect such as upturn to a steep collisional tail. With the present data, however, there no clear sign of an upturn to steeper slopes observed on P&C for $D < 1$ km (or impactors ~0.1 km in diameter), though we recognize that changes in SFD slope may be masked by a combination of sun angle, resolution, and signal-to-noise effects at these smaller sizes, even for VP. Nor is it clear that the asteroid bump at ~3–4 km (*15, 16*) and the P&C SFD slope break at 1–2 km in impactor diameter represent the same physical wave phenomenon. The strength-gravity transition is likely different for asteroids and KBOs and thus it is not clear at what KBO diameters a "wave" should occur if collisional equilibrium was reached in the outer solar system. The distinct break to the shallow slope ($q \sim -1.5$) seen on P&C, and which continues over a decade of crater/impactor sizes, is not currently produced by models of asteroid belt evolution.

Ancient state of the projectile population.

The 1-to-70-km scale craters seen on Charon's VP are below saturation at all scales and correspond to roughly 4 Gyr of bombardment (*4, 5, 28*). The relation between the crater densities and the number of projectiles in the Kuiper Belt is a complex function of the projectile orbital distribution (which affects impact probability and impact speeds and hence crater diameters) and the duration of bombardment, including the potential depopulation of projectiles over the age of the Solar System. This analysis has been performed (*4, 5*), taking into account the additional realization that some Kuiper Belt orbits cannot intersect and thus result in craters on the Pluto-Charon system. Linking the telescopically observable objects to the P&C craters is thus feasible.

Prior to the flyby, a knee Kuiper Belt model was used to predict the expected crater densities (*4, 5*). In this model the $q$ steeper than −4 slope of $d > 100$ km projectiles ($D$ ~500 km craters on Charon) shallows to $q \sim -3.5$ for smaller projectiles and was extrapolated (because there was no other information) to the smallest sizes.

The P&C crater densities for 10–100 km craters fall within the uncertainties of these previous knee model predictions. However, as described above, we find a shallower slope for crater $D \lesssim 13$ km (projectile $d \lesssim 1$–2 km), i.e., a $q$ shallower than −2. Could this shallower slope be a feature that has been preserved in the Kuiper Belt for the ~4 Gyr age of the Solar System? If so, then the SFD at this size scale cannot be greatly modified by collisions.

We can estimate the approximate collision rate as $r_c = n_V \sigma \upsilon$, where $n_V$ is the volume number density of 1-km objects, $\sigma$ the cross-sectional collision area, and $\upsilon$ the collision speed. If we take the Kuiper Belt to extend from 35 to 50 astronomical units (AU) with a scale height of 5° or ~ 0.1 radian, this yields a total vertical thickness at 40 AU of $2 \times 0.1 \times 40 = 8$ AU. This leads to an annulus of volume $V = 8 \times (50^2 - 35^2)\ \mathrm{AU}^2 = 3 \times 10^{28}\ \mathrm{km}^3$. Extrapolating the knee to $D = 1$ km ($q = -3.5$) yields $3 \times 10^9$ objects in the Kuiper belt. The cross-sectional collision area for two objects colliding is $\pi \times (\mathrm{radius}_1 + \mathrm{radius}_2)^2$ and for 1-km diameter objects the result is $\pi$ km$^2$. Lastly, colliding KBOs have an impact speed of $\upsilon^2 = \upsilon_k^2 \times (e^2 + i^2)$, where $\upsilon_k$ is the 4.5 km s$^{-1}$ Keplerian orbital speed at 43 AU (the center of the belt) and $e$ and $i$ (both about 0.05) are the orbital eccentricity and inclination in radians, yielding 300 m s$^{-1}$. This resulting collision rate is $r_c \approx (3 \times 10^9)/(3 \times 10^{28}\ \mathrm{km}^3) \times \pi\ \mathrm{km}^2 \times 0.3\ \mathrm{km\ s}^{-1} \approx 9 \times 10^{-20}\ \mathrm{s}^{-1} \sim 3\ 1 \times 1^{-12}\ \mathrm{year}^{-1} \sim 0.01$ over 4 Gyr. The estimate reflects the larger cross-sections for impactors of 1 km objects with larger but less numerous bodies, as well as the disruptive collisions with slightly smaller bodies. The estimate does not directly address the contribution of sub-km projectiles, but this adds at most a factor of a few to the disruption rate.

Overall, we find that only a few percent of multi-km objects will have suffered disruptive collisions in the last 4 Gyr. This estimate does not address the KBO accretional environment or the creation/emplacement of the Kuiper belt during giant planet migration early in Solar System history. But under the assumption that Pluto and Charon's surfaces were set early in the solar system, the projectile size distribution near the elbow has not been modified. Below that size, a $q = -2$ population has its cumulative numbers increase only as $1/d$ while the cross section drops as $d^2$ and thus smaller objects are even less collisional. A scenario in which an essentially unchanging projectile size distribution has been bombarding the Pluto-Charon system for 4 Gyr is plausible, and this distribution must be generated by accretional and emplacement models of the early Solar System (*70-73*).

## Supplementary Text

Geologic process overview.

A detailed and quantitative description of many geologic processes as applied to varied planetary conditions can be found in (*74*). Here we briefly describe each process numbered in the main text and discuss relevance to the Pluto system where possible.

1. Viscous relaxation is a wavelength dependent processes in which the creep of solid material flows under pressure gradients, such as that created by an impact crater, and tends to return material to an equipotential level (e.g., *31, 75, 76, 77*). Thus, depending on heat flow, timescales, material properties, and gravity, both negative and positive topography can be effectively flattened over time. This process has a characteristic morphologic signature: the rims of craters are retained for much longer timescales than the longer-wavelength crater bowl. No unequivocal signature of crater relaxation has been identified on either Pluto or Charon.
2. Tectonic processes on icy bodies occur in a similar manner to those on rocky bodies, where applied stress can fracture and fault material (*78*). Extensional normal faults typical of icy bodies in the outer solar system (e.g., the moons of Jupiter and Saturn) are found on both Charon and Pluto (*28, 79*), but they do not cover a large surface area and are only seen in a few cases to intersect craters.
3. Mass wasting involves down-slope movement of material such as in slumps or landslides. A few sites of large mass movement have been identified on Charon, extending from the large tectonic scarps and also potentially in the floors of larger craters (*80*), but no similarly large mass movements have been observed on Pluto. The sites in question on Charon cover a very small area.
4. Glacial flow of solid-state nitrogen ice on the surface of Pluto can carve topography and deposit sediment (*28, 81*). Glacial veneers can also shield deeper crustal water ice from smaller impacts. These processes only apply to Pluto, where the volatile ices are present on the surface. Nitrogen ice glacial flow is seen on the eastern edge of Sputnik Planitia (SP), where it appears that nitrogen ice ponded at higher altitudes has flowed into SP. There are also several types of dissected or valley-like terrains on Pluto whose morphology is likely the result of glacial erosion (*81*). There are craters in some of these terrains. Some are modified (but not erased) and some are fresher and unmodified. Today, glacial flow occurs primarily near SP.

5. Mantles of deposited material could occur on Pluto through sublimation and redisposition of volatile ices ($N_2$, CO, $CH_4$) from warmer to colder areas on seasonal cycles or million-year obliquity/precession cycles (e.g., *82, 83*). Mantles could also build up from processing of atmospheric constituents ($CH_4$, $N_2$) into larger, involatile molecules (tholin analogues), which subsequently fall out onto the surface.
6. Sublimation erosion occurs through the removal of near-surface volatile material over time due to heating by sunlight or other energetic radiation (*31*).
7. Convective activity is used here to refer to the transport of material in solid state upwellings and downwellings driven by thermally induced density contrasts. This process is thought to create the cellular/polygonal features in the Sputnik Planitia nitrogen ice sheet (*32*).
8. Icy volcanic or cryovolcanic features result from extrusion of subsurface material onto the surface (and often further horizontal flow) resulting in a partially or completely resurfaced areas. In the case of both icy convective or volcanic features on Pluto or Charon, the material may be mobile in its solid state, or partially molten (*84*).

As described in the main text, not all of these processes operate to the same extent and only mantling (process #5) appears to be preferentially erasing small craters, and then only in one major region of Pluto today (near the northern pole). Wholesale erasure of craters of all sizes by glacial erosion (process #4) may have affected specific terrains in the past (e.g., eastern Tombaugh Regio).

We also comment on resurfacing by the craters themselves. The formation of the basin that encloses Sputnik Planitia on Pluto was likely a very large impact and thus a major resurfacing event (including a regionally extensive ejecta deposit). The basin appears very ancient (*85*), however, in this work we assume that all Pluto surfaces mapped post-date basin formation (nor do we include Sputnik "basin" in the R-plots in Fig. 2A). The ejecta deposits of mid-sized and smaller craters on Pluto and Charon (*86*) can resurface locally. At the resolution of the New Horizons images, however, there are only a few obvious ejecta deposits (mostly on Charon) and no clear cases of partially filled craters by these ejecta deposits. This is consistent with the ejecta being a relatively thin deposit and thus not able to fill or partially-fill many craters of the scale we are examining.

<u>Largest craters on Pluto and Charon and overlap with telescopic survey data.</u>

Few telescopic surveys have probed KBOs as small as those that formed craters on Pluto and Charon. A handful of surveys have observed objects below the knee break in slope ($H_g$ ~ 9) and fainter objects, down to $H_g$ ~12 (*18-20, 40-43, 87, 88*), which equates to KBO diameters down to ~20 km for an assumed $H_g$ albedo of 5%. The largest four craters on Pluto's encounter hemisphere that overlap with directly observed KBOs include: (1) Sputnik basin, the likely impact structure underlying Sputnik Planitia, for which the equivalent rim location could be in the range of $D$ ~ 600–1000 km ($d$ ~ 150–275 km); (2) Burney, a multi-ring basin, at $D \approx$ 240 km ($d \approx$ 50 km; **fig. S3c**); (3) Edgeworth crater at $D \approx$ 145 km ($d \approx$ 28 km); and (4) Oort crater at $D \approx$ 120 km ($d$ ~ 22 km). Sputnik basin likely represents the only basin of its size on Pluto, as no similar albedo features are seen and would likely be visible even in the lower-resolution images of the non-encounter hemisphere of Pluto. Oort and Edgeworth are both located in the Cthulhu region of Pluto, the most heavily cratered (at larger $D$) and thus the most ancient region on Pluto's encounter hemisphere. On Charon, only the impactor that created the largest crater observed on the encounter hemisphere, Dorothy (Gale) at $D \approx 240$ km ($d \sim 37$ km), overlaps with the smallest observed KBOs from the above surveys.

Occultation surveys have probed for even smaller objects, down to $d \sim 0.5$ km with ~25 total candidate detections from two surveys (*46, 47, 89, 90*), and no detections in a third survey (*91*). These surveys calculated some constraints on the slope of the KBO SFD for $d < 100$ km. The survey with no detections (*91*) finds an upper limit $q$ of 3.34–3.82 for objects between ~90 km in diameter and the detection limit estimated to be ~0.5 km in diameter. This result is consistent with the SFD slope found for the Pluto and Charon craters above the "elbow" ($D > 10$ km or $d > 1$–2 km), but would be too steep compared to our findings for KBOs below the "elbow" ($d < 1$–2 km). The surveys with candidate detections (*46, 47, 89, 90*) and the collisional evolution model that matches them (*3*) estimate steeper slopes for KBOs below the "elbow" ($q \sim$ 3–4.5), and also estimate the density of smaller KBOs to be much higher than is suggested by the crater data for Pluto and Charon (e.g., *3*) (see main text for additional discussion).

Implications for Comet Collisional Evolution

An ongoing debate concerns whether comets could survive into the present as primordial objects, or if they are products of collisions and other processes that re-form them over the course of solar system history (*11-13, 92, 93*). These two alternate histories for comets carry different implications for interpreting their structure and composition. The bi-lobed comet 67P/Churyumov–Gerasimenko (67P/CG) is ~ 7 km by 4 km across in cross section (*94*), larger than the elbow SFD slope break. Dynamical simulations of 67P/CG's collisional history depend on the number of comets (as derived from the total mass and assumed SFD shape). A shallower slope assumption motivated by the New Horizons results leads to a higher chance of 67P/CG surviving intact into the present as calculated by some models (*12*).

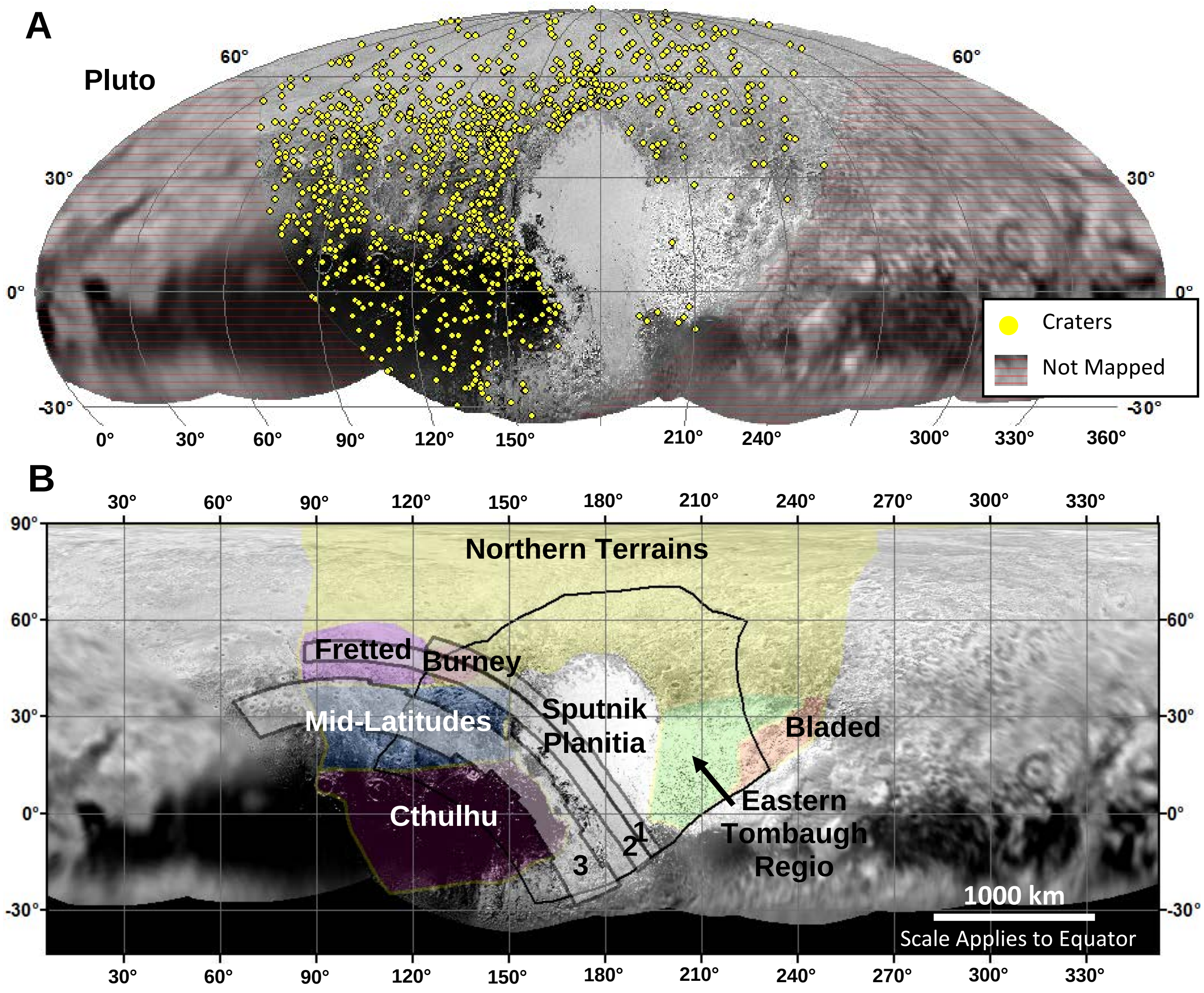

**Fig. S1: Mapped regions on Pluto.**

(**A**) Equal-area Mollweide projection of craters mapped on the PELR_P_LORRI mosaic (850 m px$^{-1}$). Background mosaic contains images from the PELR_P_LORRI sequence as well as PELR_P_LORRI_STEREO_MOSAIC (~400 m px$^{-1}$) and additional lower resolution LORRI images covering the non-encounter hemisphere. (**B**) Simple cylindrical projection of the same mosaic with broad regions delineated. Colored areas indicate extent of mappable PEMV_P_MPAN1 (480 m px$^{-1}$) and solid black outline encompasses the more limited area from PEMV_P_MVIC_LORRI_CA (315 m px$^{-1}$). The high resolution strips are also shown and labelled 1, 2, and 3 for PELR_P_MVIC_LORRI_CA (76 m px$^{-1}$), PELR_P_MPAN_1 (117 m px$^{-1}$), and PELR_P_LEISA_HiRES (234 m px$^{-1}$), respectively. See Fig. 2 for Crater SFDs.

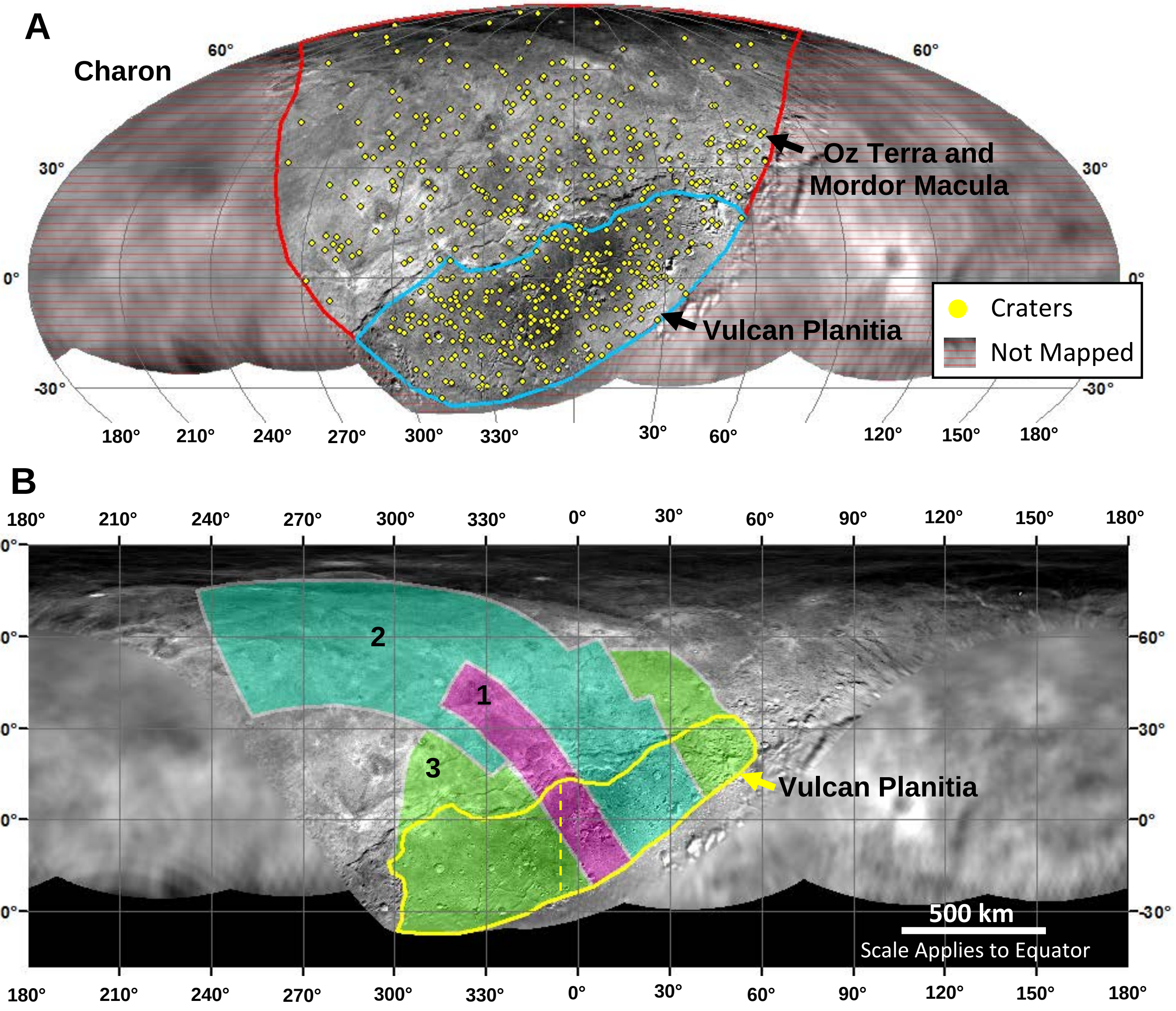

**Fig. S2: Mapped regions on Charon.**

(**A**) Equal-area Mollweide projection of craters mapped on the PELR_C_LORRI (865 m $px^{-1}$) mosaic. Red and blue outlines indicate the mappable extent of Oz Terra and Vulcan Planitia, respectively. (**B**) Simple cylindrical projection showing extent of mappable areas (limited by each mosaic's coverage and lighting geometry) of PELR_C_MVIC_LORRI_CA (labelled "1"; 154 m $px^{-1}$), PELR_C_LEISA_HIRES (labelled "2"; 410 m $px^{-1}$), and PEMV_C_MVIC_LORRI_CA (labelled "3"; 622 m $px^{-1}$). Yellow outline indicates the mappable extent of VP in PEMV_C_MVIC_LORRI_CA. Dashed yellow line indicates center line splitting VP into approximately equal-area east and west halves. See Fig. 2 for Crater SFDs.

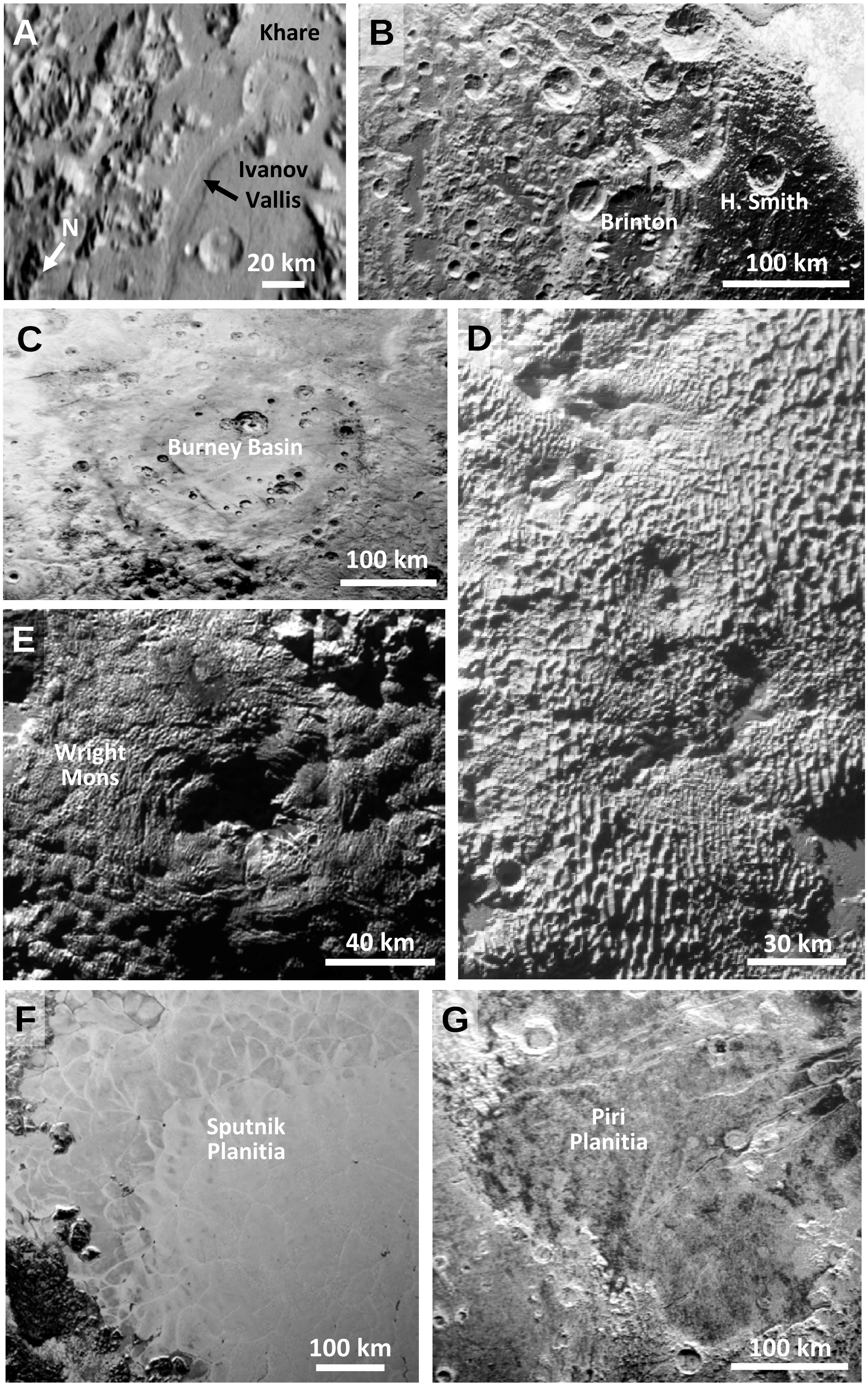

**Fig. S3: Regions on Pluto discussed in the paper.**

These include: (**A**) mantled terrains near Pluto's north pole, (**B**) heavily cratered portion of Cthulhu Regio, (**C**) Burney, a multi-ring basin, and the largest crater after Sputnik basin on Pluto's encounter hemisphere, (**D**) bladed terrain in the eastern portion of Pluto's encounter hemisphere, (**E**) one of Pluto's putative cryovolcanic constructs, Wright Mons (**F**) the vast $N_2$-$CH_4$-CO ice sheet of Sputnik Planitia with no identifiable craters, (**G**) a region of low crater density that appears to have been resurfaced by scarp retreat. See (*27, 28, 60, 95*) for more description of these regions. Image request IDs and resolutions: (A,C, D, G) PEMV_P_MPAN1 (480 m $px^{-1}$); (B, E, F) PEMV_P_MVIC_LORRI_CA (315 m $px^{-1}$). Images are north up.

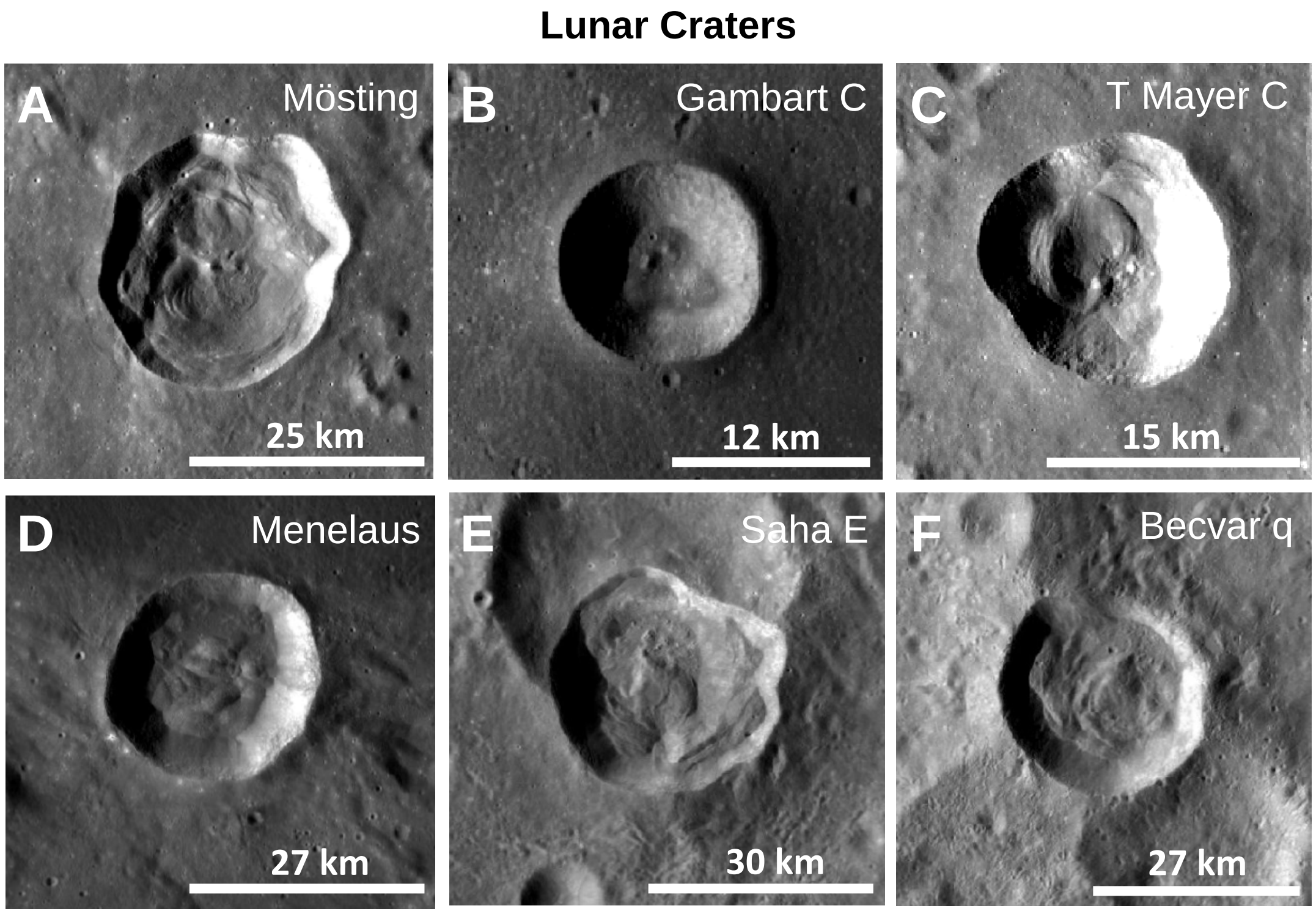

**Fig. S4: Lunar craters with hummocky floors.**

(**A-C**) Mare, (**D**) partial mare/partial highlands, and (**E–F**) highlands examples**.** These craters display mass wasting and terrace slumps, both of which create various hummocky features on the crater floors. These may be analogous to some of the hummocky crater floors observed on Charon. Images are taken from the 100 m $px^{-1}$ Lunar Reconnaissance Orbiter Camera (LROC) wide angle camera basemap (*96*).

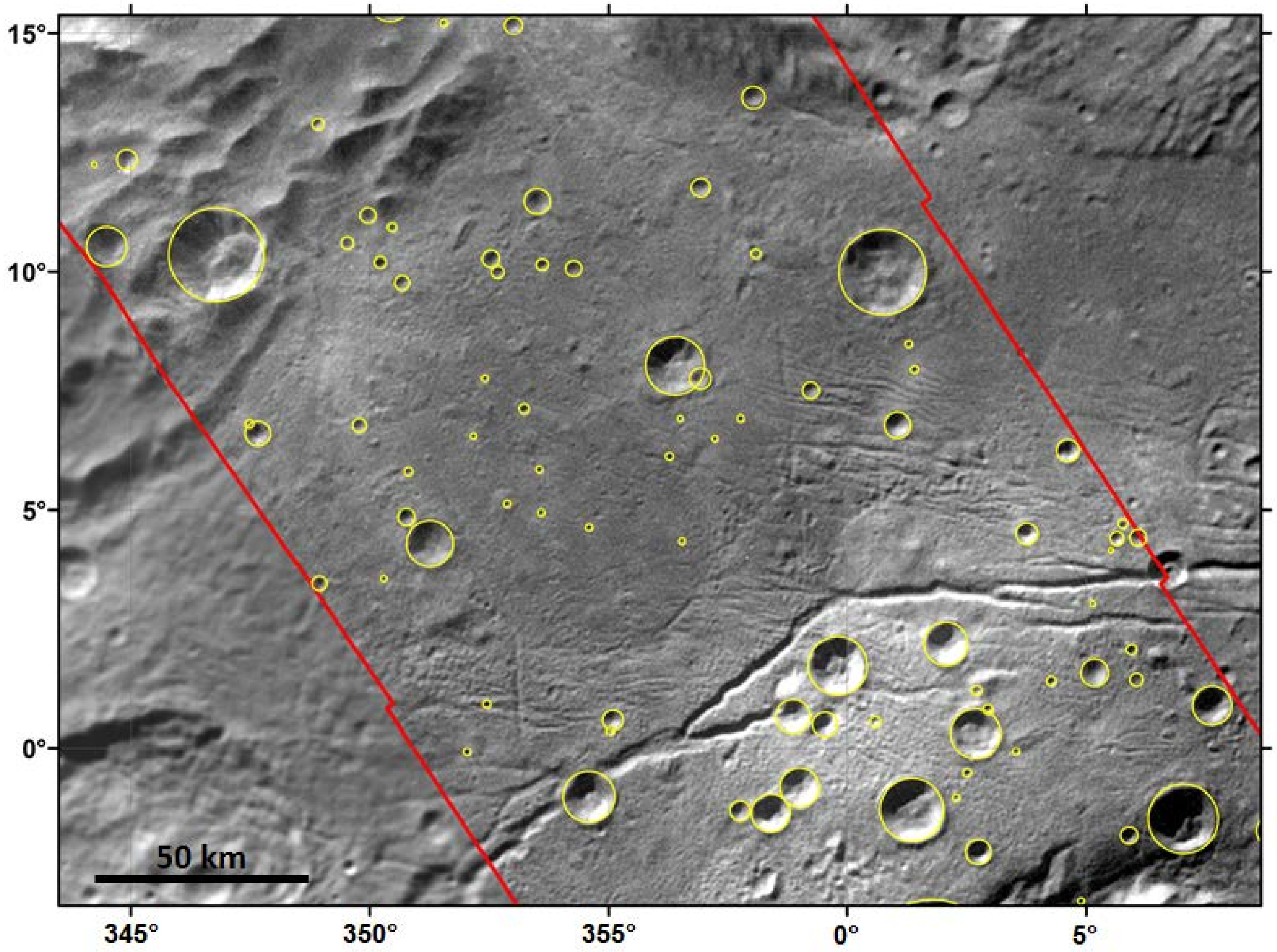

**Fig S5: Crater distribution on Charon's Vulcan Planitia at high resolution.**
Yellow circles indicate mapped craters on the smooth plains region of Vulcan Planitia covered by the highest resolution image strip (154 m px$^{-1}$ PELR_C_MVIC_LORRI_CA, red outline). Although all mapped craters are shown here, only craters larger than 1.4 km (~10 pixels) are included in the R-plots and the slope calculations in Table S2. Several small depressions that were not mapped as craters can be seen. These features were considered to be more likely part of ridge or trough systems (or other topography) rather than a crater because they lack distinct raised rims and/or were not very circular in planform.

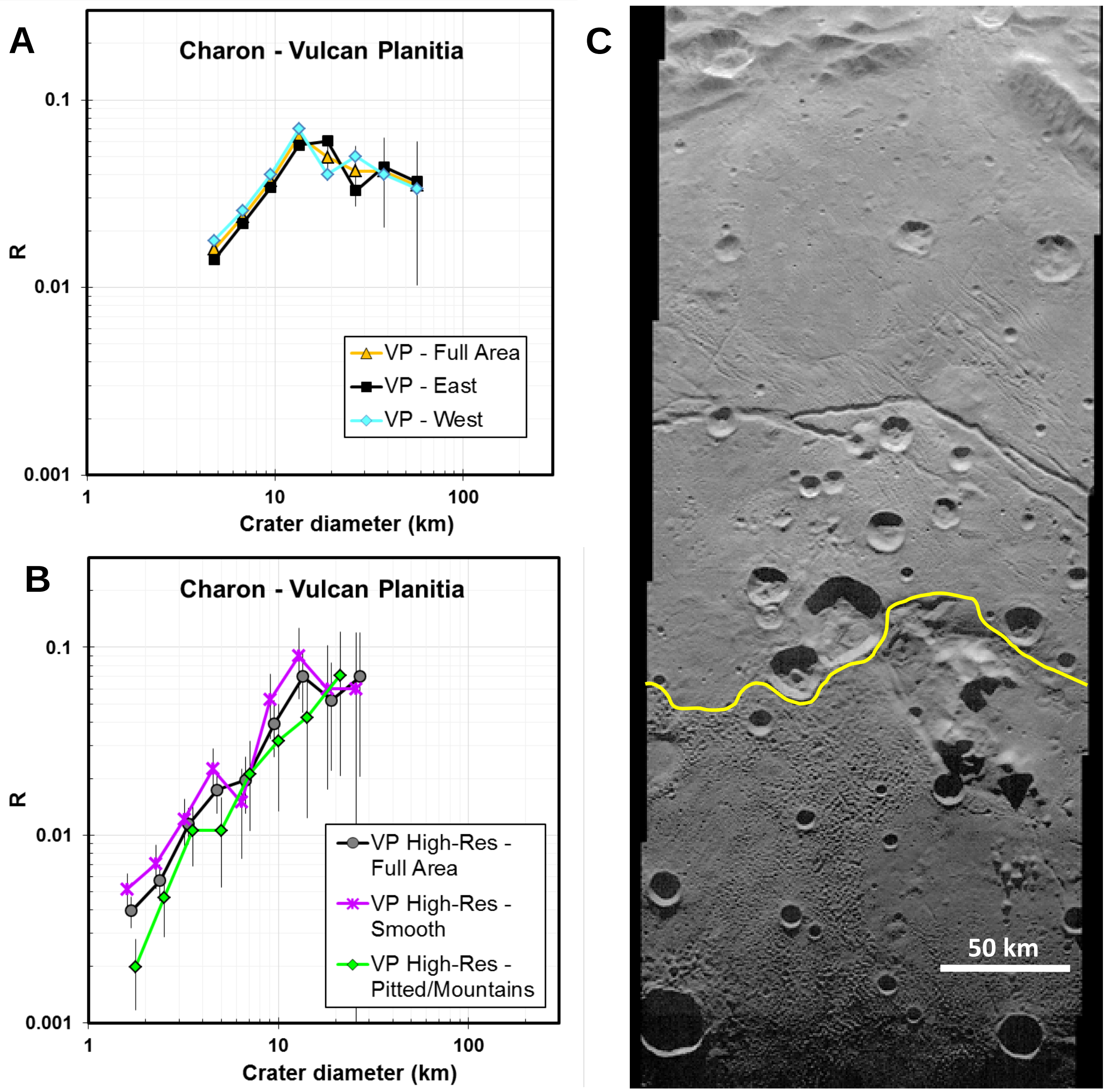

**Fig. S6: Crater SFDs for different portions of Vulcan Planitia.**
(**A**) Vulcan Planitia is arbitrarily split into two halves (east and west) at −5° of longitude (**see fig. S2**). Data from the 622 m px$^{-1}$ PEMV_C_MVIC_LORRI_CA image scan ("Full Area" is also shown in Figs. 2 and 3). Both regions share a similar SFD to the entire plain, providing further evidence that the SFD shape is likely a reflection of the exogenous impactor flux, rather than icy volcanism alone erasing smaller craters. Volcanism acted on Charon but it would be unlikely to have equally removed the same number of small craters in the same diameter range in all areas across VP. In addition, there are no obvious partially filled craters on VP itself (see discussion in text). (**B-C**) Craters in the highest resolution strip (154 m px$^{-1}$ PELR_C_MVIC_LORRI_CA), split into the northern area of smoother terrains, and the southern area with textured terrain and "mountains in moats". The "Smooth" and "Pitted/Mountains" distributions have been shifted by 5% from the "Full Area" to make the error bars distinguishable. The largest texture elements (pits and knobs) in the southern area have a scale of ~4 km, but the characteristic texture scale is closer to 2 km. The distributions are similar but the textured/pitted terrain (26 craters) has a somewhat shallower slope overall (30 craters; $q$ = –1.4 weighted differential) than the smooth terrain region (70 craters; $q$ = –1.7). We use the slope of the smooth terrain in Table 2 as a conservative measure, to minimize confusion between impact craters and endogenic texture. We do include the pitted terrain for reference in Data S1–S3but we recommend a higher diameter cutoff for this dataset ($D$ > 4 km) due to the background terrain texture and near-terminator lighting. Panel C extends from 345° to 17° in longitude and 17° to −21° in latitude (also see figs. S2, S5).

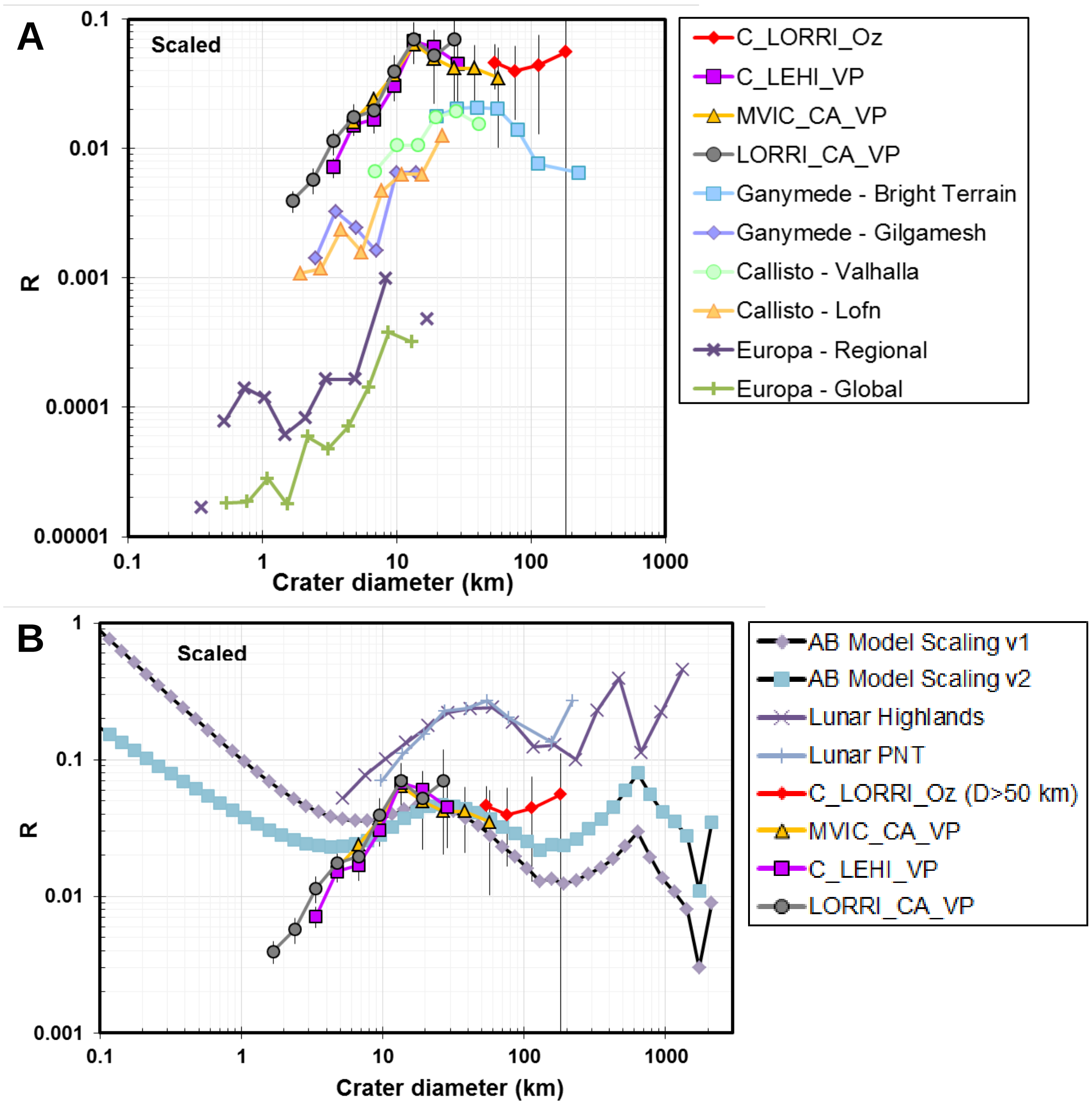

**Fig S7: Charon crater SFDs compared to other cratered worlds.**
(**A**) Young terrains on Jupiter's moons display similarly shallow crater SFD slopes to those seen on Charon over a similar scaled crater diameter range (*6*, *8*, *36*). Craters on Neptune's moon

Triton, however, have steeper SFD slopes in this diameter range (*6, 97*). (**B**) The SFD shape of craters on the Moon (e.g., *98, 99, 100*) as well as that of a model of collisional evolution of the asteroid belt (*15, 16*) differs from that of Charon at the smaller sizes ($D$ < 10 km). Note change in y-axis scale between plots. See methods for scaling and table S3 for description of terrains.

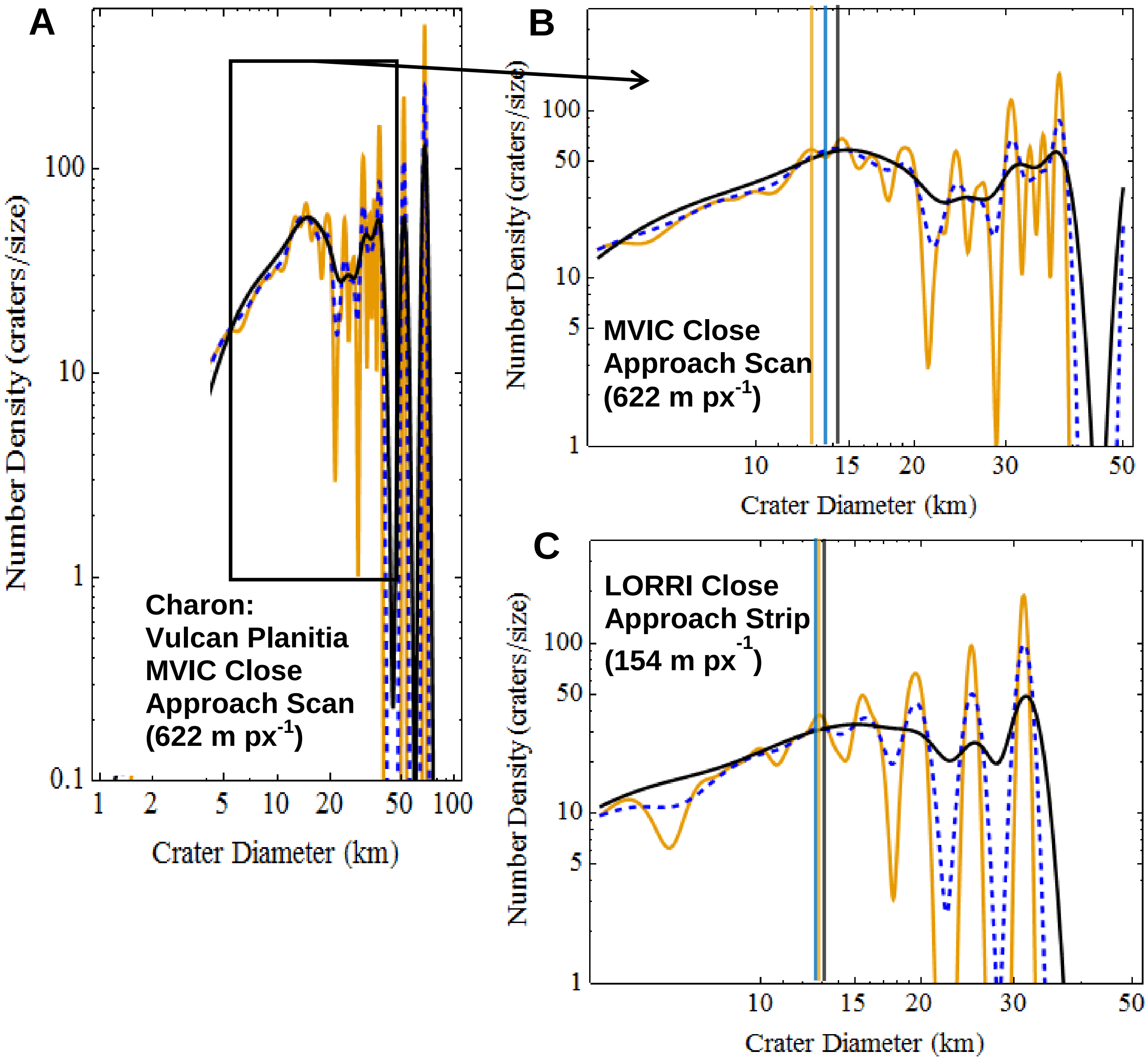

**Fig. S8: Characterizing the crater SFD slope break for Vulcan Planitia.** Smooth kernel distribution R-plots (Gaussian kernel) with varying bandwidths for several datasets covering Charon's Vulcan Planitia: (**A–B**) PEMV_C_MVIC_LORRI_CA (622 m px$^{-1}$), and (**C**) PELR_C_MVIC_LORRI_CA (154 m px$^{-1}$). Although larger bandwidths cause a slight shift in the break location toward larger $D$, the break location appears to be around 13 km in diameter (vertical lines). Note change in plot aspect ratio for B and C.

**Table S1. Image datasets and mosaics.** PELR refers to a "Pluto Encounter LORRI" observation, and PEMV refers to a "Pluto Encounter MVIC" observation. Refer to (*55, 57*) for specifics on the LORRI and MVIC instruments. Pixel scale varied across the scene in each case as the spacecraft moved closer to the body during the observations. The values here are the approximate average pixel scale covering the majority of the mapped surface area with an estimate of the variation based on the changing spacecraft distance. We did not map in any areas near the limb in a given image set (areas observed extremely obliquely). If the pixel scale variation is not listed, it is 1 m $px^{-1}$ or less. Higher resolution strips (indicated with ‡) were taken as sequential ride-along LORRI frames during scanning of another instrument. For the ride-along observations, the scan size only includes the images that fall on the illuminated surface of either body.

| Request ID | Plot Legend Title | Instrument Mode | ~Pixel Scale [m $px^{-1}$] | Mosaic size or Scan | Exposure or Scan rate |
|---|---|---|---|---|---|
| Pluto | | | | | |
| PELR_P_LORRI | | 1×1 | 850 ± 30 | 4×5 | 150 ms |
| PELR_P_LEISA_HIRES‡ | | 1×1 | 234 ± 13 | 1×12 | 50 ms |
| PELR_P_MPAN_1‡ | | 1×1 | 117 ± 2 | 1×27 | 10 ms |
| PELR_P_ MVIC_ LORRI_CA‡ | | 1×1 | 76 | 1x35 | 10 ms |
| PEMV_P_MPAN1 | MPAN | Pan TDI 1 | 480 ± 5 | Scan | 1600 μrad $s^{-1}$ |
| PEMV_P_ MVIC_ LORRI_CA | MCA | Pan TDI 2 | 315 ± 8 | Scan | 1000 μrad $s^{-1}$ |
| | | | | | |
| Charon | | | | | |
| PELR_C_LORRI | C_LORRI | 1×1 | 865 ± 10 | 2×4 | 150 ms |
| PELR_C_LEISA_HIRES‡ | C_LEHI | 1×1 | 410 ± 5 | 1×7 | 60 ms |
| PELR_C_ MVIC_ LORRI_CA‡ | LORRI_CA | 1×1 | 154 | 1×9 | 10 ms |
| PEMV_C_ MVIC_ LORRI_CA | MVIC_CA | Pan TDI 1 | 622 | Scan | 1000 μrad $s^{-1}$ |

**Table S2. Pluto and Charon crater size−frequency distribution slopes.** Differential slopes are calculated by linear regression on the log of the data. Uncertainties are the standard least-squared errors of the parameters (one sigma). Weighted regressions use *N* (the variance), the number of craters in each bin, as the weights. Maximum likelihood slopes are estimated using a truncated Pareto distribution (*101, 102*) in Mathematica. Uncertainties are standard one-sigma errors (*101, 102*). Positive slopes cannot be calculated with the Pareto distribution (marked as N/A). The cumulative distribution slope is obtained by adding 1 to the differential slope listed here (e.g, the cumulative slope for the first row is −1.7). For the craters *D* < 10 km in the highest resolution image set, we used a subset of the Vulcan Planitia area. See figs. S5, S6, and their figure captions for description.

| Terrain | Crater Diameter Range (km) | Number of Craters | Un−weighted Differential Slope | Weighted Differential Slope | Maximum Likelihood Differential Slope |
|---|---|---|---|---|---|
| Pluto | | | | | |
| MPAN_Cthulhu | >10 | 186 | −2.7 ± 0.1 | −2.7 ± 0.1 | −2.7 ± 0.1 |
| | 3–10 | 337 | −0.6 ± 0.2 | −0.7 ± 0.2 | N/A |
| MPAN_Mid−lat | >15 | 69 | −3.8 ± 0.1 | −3.7 ± 0.1 | −3.6 ± 0.4 |
| | 3–15 | 138 | 0.1 ± 0.4 | 0.2 ± 0.5 | N/A |
| MPAN_Burney | >10 | 12 | −2.8 ± 0.7 | −3.0 ± 0.6 | −2.3 ± 1.0 |
| | 3–10 | 64 | −1.8 ± 0.5 | −1.7 ± 0.5 | −1.6 ± 0.4 |
| MPAN_Fretted | >10 | 56 | −2.6 ± 0.5 | −2.2 ± 0.5 | −2.5 ± 0.3 |
| | 3–10 | 65 | 0.1 ± 0.2 | 0.1 ± 0.2 | N/A |
| MPAN_North | >10 | 167 | −3.1 ± 0.2 | −2.8 ± 0.2 | −2.6 ± 0.2 |
| | 3–10 | 161 | 0.2 ± 0.2 | 0.2 ± 0.3 | N/A |
| MCA_Cthulhu | >10 | 76 | −2.3 ± 0.3 | −2.6 ± 0.3 | −2.9 ± 0.3 |
| | 2–10 | 214 | −0.7 ± 0.1 | −0.7 ± 0.1 | N/A |
| MCA_Mid−lat | >15 | 44 | −3.5 ± 0.1 | −3.4 ± 0.1 | −3.0 ± 0.5 |
| | 2–15 | 127 | 0.9 ± 0.5 | 0.0 ± 0.5 | N/A |
| MCA_Burney | >10 | 13 | −2.1 ± 0.2 | −2.1 ± 0.2 | −2.3 ± 0.8 |
| | 2–10 | 54 | −0.8 ± 0.5 | −0.8 ± 0.5 | N/A |
| Charon | | | | | |
| C_LORRI_Oz | >50 | 13 | −2.8 ± 0.1 | −2.9 ± 0.2 | −3.3 ± 0.8 |
| MVIC_CA_VP | >10 | 97 | −3.4 ± 0.1 | −3.5 ± 0.1 | −3.1 ± 0.3 |
| | 4–10 | 213 | −1.8 ± 0.04 | −1.8 ± 0.04 | −1.7 ± 0.3 |
| C_LEHI_VP | >10 | 35 | −3.5 ± 0.1 | −3.5 ± 0.1 | −2.7 ± 0.6 |
| | 3–10 | 100 | −1.7 ± 0.2 | −1.7 ± 0.3 | −1.3 ± 0.3 |
| LORRI_CA_VP | >10 | 16 | −3.0 ± 0.5 | −3.2 ± 0.4 | −2.8 ± 0.8 |
| (smooth area)§ | 1.4–10 | 70 | −1.8 ± 0.2 | −1.7 ± 0.2 | −1.8 ± 0.2 |

**Table S3. Crater data information for Figure S6.** Multiply the crater size on these bodies by the scale factor (SF) to arrive at the equivalent size crater the same impactor would make on Charon (*1*).

| Panel | Name in Figure S6 Legend | Average Impact Velocity (*6, 103, 104*) [km s$^{-1}$] | SF | Description | Source |
|---|---|---|---|---|---|
| **a** | Europa - Regional | 26 | 0.46 | Areas covered by Galileo's regional mapping campaign (~10% of the surface at ~220 m px$^{-1}$) | (*36*) |
| **a** | Europa - Global | 26 | 0.46 | ~100 craters with $D > 1$ km – mapped where data permits | (*6*, *8*) |
| **a** | Ganymede - Bright Terrain | 20 | 0.52 | All craters on bright terrain where observable. | (*8*) |
| **a** | Ganymede - Gilgamesh | 20 | 0.52 | Craters on Ganymede's youngest impact basin Gilgamesh | (*8*) |
| **a** | Callisto -Valhalla | 15 | 0.57 | Craters on the relatively young impact basin Valhalla | (*8*) |
| **a** | Callisto - Lofn | 15 | 0.57 | Craters on the relatively young impact basin Lofn | (*8*) |
| **b** | AB Model Scaling v1, v2 | | | Model of asteroid belt collisional evolution. (see methods for scaling info) | (*15*, *16*) |
| **b** | Lunar Highlands | 18.9 | 0.55 | Front-side highlands | (*98*, *105*) |
| **b** | Lunar PNT | 18.9 | 0.55 | Lunar Pre-Nectarian-era terrains | (*99*) |

**Data S1 (separate file)**

This file contains a list of all craters for Pluto and Charon presented here. See methods section and tables S1 and S2 for information on feature selection, mapping methods, and dataset properties. This additional file contains a list of the crater diameters (in km), the crater locations (latitude and longitude in degrees), the dataset name for the image dataset the craters are derived from (table S1), the name for the dataset and geologic region, and in some cases an additional designation for a subset of the geologic region (as described in the text, shown in figs. S1-S2, and S5 and table S2). This file contains all identified down to the resolution limit, including some craters smaller than were included in our analysis. For diameter cutoffs used in the analysis of the SFD slopes see table S2.

**Data S2 (separate file)**

This file contains a list of the surface areas in km$^2$ for each geologic region and dataset combination described in the text and given in table S2.

**Data S3 (separate file)**

This file contains a list of the R-value data points for Pluto and Charon as displayed in the main text and supplementary figures.

1. Materials and methods are available in the supplementary material.
2. C. A. T. W. Group *et al.*, Standard techniques for presentation and analysis of crater size-frequency data. *Icarus* **37**, 467-474 (1979). doi: 10.1016/0019-1035(79)90009-5
3. H. E. Schlichting, C. I. Fuentes, D. E. Trilling, Initial planetesimal sizes and the size distribution of small Kuiper belt objects. *Astron. J.* **146**, 36 (2013).
4. S. Greenstreet, B. Gladman, W. B. McKinnon, Corrigendum to "impact and cratering rates onto Pluto" [icarus 258 (2015) 267-288]. *Icarus* **274**, 366-367 (2016). doi: 10.1016/j.icarus.2016.03.003
5. S. Greenstreet, B. Gladman, W. B. McKinnon, Impact and cratering rates onto Pluto. *Icarus* **258**, 267-288 (2015). doi: 10.1016/j.icarus.2015.05.026
6. K. Zahnle, P. Schenk, H. Levison, L. Dones, Cratering rates in the outer Solar System. *Icarus* **163**, 263-289 (2003).
7. A. Morbidelli *et al.*, in *The Solar System beyond Neptune,* M. A. Barucci, Ed. (2008), pp. 275-292.
8. P. M. Schenk, C. R. Chapman, K. Zahnle, J. M. Moore, in *Jupiter*, F. Bagenal, T. E. Dowling, W. B. McKinnon, Eds. (Cambridge Univ. Press, New York, 2004), pp. 427-456.
9. L. Dones *et al.*, in *Saturn from Cassini-Huygens,* M. K. Dougherty, Ed. (2009), pp. 613-635.
10. M. Kirchoff *et al.*, in *Enceladus and the icy moons of Saturn,* P. Schenk, C. Howett, Eds. (University of Arizona Press, 2018), pp. 267-306.

11. B. J. R. Davidsson *et al.*, The primordial nucleus of comet 67P/churyumov-gerasimenko. *A & A* **592**, (2016). doi: 10.1051/0004-6361/201526968
12. M. Jutzi, W. Benz, A. Toliou, A. Morbidelli, R. Brasser, How primordial is the structure of comet 67P?. Combined collisional and dynamical models suggest a late formation. *A & A* **597**, (2017). doi: 10.1051/0004-6361/201628963
13. A. Morbidelli, H. Rickman, Comets as collisional fragments of a primordial planetesimal disk. *A & A* **583**, (2015). doi: 10.1051/0004-6361/201526116
14. E. B. Bierhaus, L. Dones, Craters and ejecta on Pluto and Charon: Anticipated results from the New Horizons flyby. *Icarus* **246**, 165-182 (2015). doi: 10.1016/j.icarus.2014.05.044
15. W. F. Bottke *et al.*, The fossilized size distribution of the main asteroid belt. *Icarus* **175**, 111-140 (2005).
16. W. F. Bottke *et al.*, Linking the collisional history of the main asteroid belt to its dynamical excitation and depletion. *Icarus* **179**, 63-94 (2005). doi: 10.1016/j.icarus.2005.05.017
17. M. J. Lehner *et al.*, in *SPIE Astronomical Telescopes + Instrumentation*. (SPIE, 2016), vol. 9906, pp. 10.
18. G. M. Bernstein *et al.*, The size distribution of trans-neptunian bodies. *Astron. J.* **128**, 1364-1390 (2004). doi: 10.1086/422919
19. W. C. Fraser, M. E. Brown, A. Morbidelli, A. Parker, K. Batygin, The absolute magnitude distribution of Kuiper belt objects. *Astrophys J.* **782**, (2014). doi: 10.1088/0004-637X/782/2/100
20. M. Alexandersen *et al.*, A carefully characterized and tracked trans-neptunian survey: The size distribution of the plutinos and the number of neptunian trojans. *Astron. J.* **152**, (2016).
21. M. W. Buie *et al.*, paper presented at the 49th Annual American Astronomical Society Division for Planetary Sciences Meeting, Provo, UT, 1 October 2017.
22. D. A. Minton, J. E. Richardson, P. Thomas, M. Kirchoff, M. E. Schwamb, paper presented at the Asteroids, Comets, Meteors 2012 Conference, Niigata, Japan, 1 May 2012, #6348.
23. E. Chiang, A. N. Youdin, Forming planetesimals in solar and extrasolar nebulae. *Annu. Rev. Earth Planet Sci.* **38**, 493-522 (2010). doi: 10.1146/annurev-earth-040809-152513
24. D. Nesvorný, A. N. Youdin, D. C. Richardson, Formation of Kuiper belt binaries by gravitational collapse. *Astron. J.* **140**, 785-793 (2010). doi: 10.1088/0004-6256/140/3/785
25. A. Johansen, M.-M. Mac Low, P. Lacerda, M. Bizzarro, Growth of asteroids, planetary embryos, and Kuiper belt objects by chondrule accretion. *Science Advances* **1**, (2015). doi: 10.1126/sciadv.1500109
26. J. B. Simon, P. J. Armitage, R. Li, A. N. Youdin, The mass and size distribution of planetesimals formed by the streaming instability. I. The role of self-gravity. *Astrophys J.* **822**, (2016).
27. S. A. Stern *et al.*, The Pluto system: Initial results from its exploration by New Horizons. *Science* **350**, id.aad1815 (2015). doi: 10.1126/science.aad1815
28. J. M. Moore *et al.*, The geology of Pluto and Charon through the eyes of New Horizons. *Science* **351**, 1284-1293 (2016). doi: 10.1126/science.aad7055
29. H. J. Melosh, *Impact cratering: A geologic perspective*. (Oxford Univ. Press, New York, 1989), pp. 253.

30. M. Gurnis, Martian cratering revisited - implications for early geologic evolution. *Icarus* **48**, 62-75 (1981). doi: 10.1016/0019-1035(81)90154-8
31. S. A. Stern, S. Porter, A. Zangari, On the roles of escape erosion and the viscous relaxation of craters on Pluto. *Icarus* **250**, 287-293 (2015). doi: 10.1016/j.icarus.2014.12.006
32. W. B. McKinnon *et al.*, Convection in a volatile nitrogen-ice-rich layer drives Pluto's geological vigour. *Nature* **534**, 82-85 (2016). doi: 10.1038/nature18289
33. S. A. Stern *et al.*, New Horizons constraints on Charon's present day atmosphere. *Icarus* **287**, 124-130 (2017). doi: 10.1016/j.icarus.2016.09.019
34. D. E. Wilhelms, "The geologic history of the moon" (1987), QB592.W5.
35. E. B. Bierhaus, C. R. Chapman, W. J. Merline, Secondary craters on Europa and implications for cratered surfaces. *Nature* **437**, 1125-1127 (2005). doi: 10.1038/nature04069
36. E. B. Bierhaus, K. Zahnle, C. R. Chapman, in *Europa*, R. T. Pappalardo, W. B. McKinnon, K. K. Khurana, Eds. (Univ. Arizona Press, Tucson, 2009), pp. 161-180.
37. A. S. McEwen, E. B. Bierhaus, The importance of secondary cratering to age constraints on planetary surfaces. *Annu. Rev. Earth Planet Sci.* **34**, 535-567 (2006). doi: 10.1146/annurev.earth.34.031405.125018
38. K. N. Singer, W. B. McKinnon, L. T. Nowicki, Secondary craters from large impacts on Europa and Ganymede: Ejecta size–velocity distributions on icy worlds, and the scaling of ejected blocks. *Icarus* **226**, 865-884 (2013). doi: 10.1016/j.icarus.2013.06.034
39. R. A. Smullen, K. M. Kratter, The fate of debris in the Pluto-Charon system. *Mon. Not. R. Astron. Soc.* **466**, 4480-4491 (2017). doi: 10.1093/mnras/stw3386
40. W. C. Fraser, J. J. Kavelaars, A derivation of the luminosity function of the Kuiper belt from a broken power-law size distribution. *Icarus* **198**, 452-458 (2008).
41. C. I. Fuentes, M. J. Holman, A SUBARU archival search for faint trans-neptunian objects. *Astron. J.* **136**, 83-97 (2008). doi: 10.1088/0004-6256/136/1/83
42. B. Gladman *et al.*, The structure of the Kuiper belt: Size distribution and radial extent. *Astron. J.* **122**, 1051-1066 (2001). doi: 10.1086/322080
43. A. H. Parker *et al.*, paper presented at the 46th Lunar and Planetary Science Conference, Houston, TX, 1 March 2015, #2614.
44. C. Shankman, B. J. Gladman, N. Kaib, J. J. Kavelaars, J. M. Petit, A possible divot in the size distribution of the Kuiper belt's scattering objects. *ApJ Letters* **764**, (2013). doi: 10.1088/2041-8205/764/1/L2
45. C. Shankman *et al.*, OSSOS. II. A sharp transition in the absolute magnitude distribution of the Kuiper belt's scattering population. *Astron. J.* **151**, (2016). doi: 10.3847/0004-6256/151/2/31
46. H. E. Schlichting *et al.*, Measuring the abundance of sub-kilometer-sized Kuiper belt objects using stellar occultations. *Astrophys J.* **761**, (2012). doi: 10.1088/0004-637X/761/2/150
47. H. E. Schlichting *et al.*, A single sub-kilometre Kuiper belt object from a stellar occultation in archival data. *Nature* **462**, 895-897 (2009). doi: 10.1038/nature08608
48. W. F. Bottke *et al.*, in *Asteroids IV*, P. Michel, Ed. (University of Arizona Press, Tucson, 2015), pp. 701-724.
49. M. Pan, R. e. Sari, Shaping the Kuiper belt size distribution by shattering large but strengthless bodies. *Icarus* **173**, 342-348 (2005).

50. S. J. Kenyon, B. C. Bromley, Coagulation calculations of icy planet formation at 15-150 au: A correlation between the maximum radius and the slope of the size distribution for trans-neptunian objects. *Astron. J.* **143**, (2012).
51. S. J. Kenyon, B. C. Bromley, The size distribution of Kuiper belt objects. *Astron. J.* **128**, 1916-1926 (2004).
52. A. Youdin, A. Johansen, Protoplanetary disk turbulence driven by the streaming instability: Linear evolution and numerical methods. *Astrophys J.* **662**, 613-626 (2007). doi: 10.1086/516729
53. J. B. Simon, P. J. Armitage, A. N. Youdin, R. Li, Evidence for universality in the initial planetesimal mass function. *ApJ Letters* **847**, (2017).
54. C. P. Abod *et al.*, in *arXiv e-prints*. (2018), vol. 1810.
55. A. F. Cheng *et al.*, Long-Range Reconnaissance Imager on New Horizons. *Space Sci. Rev.* **140**, 189-215 (2008). doi: 10.1007/s11214-007-9271-6
56. C. J. A. Howett *et al.*, Inflight radiometric calibration of New Horizons' multispectral visible imaging camera (MVIC). *Icarus* **287**, 140-151 (2017). doi: 10.1016/j.icarus.2016.12.007
57. D. C. Reuter *et al.*, Ralph: A visible/infrared imager for the New Horizons Pluto/Kuiper belt mission. *Space Sci. Rev.* **140**, 129-154 (2008). doi: 10.1007/s11214-008-9375-7
58. New Horizons PDS page. **https://pdssbn.astro.umd.edu/data_sb/missions/newhorizons/index.shtml**, (2018).
59. L. Keszthelyi *et al.*, paper presented at the Lunar Planet. Sci., Houston, TX, 1 March 2013, #2546.
60. P. M. Schenk *et al.*, Basins, fractures and volcanoes: Global cartography and topography of Pluto from New Horizons. *Icarus* **314**, 400-433 (2018). doi: 10.1016/j.icarus.2018.06.008
61. P. M. Schenk *et al.*, Breaking up is hard to do: Global cartography and topography of Pluto's mid-sized icy moon Charon from New Horizons. *Icarus* **315**, 124-145 (2018). doi: 10.1016/j.icarus.2018.06.010
62. S. J. Robbins *et al.*, Craters of the Pluto-Charon system. *Icarus* **287**, 187-206 (2017). doi: 10.1016/j.icarus.2016.09.027
63. S. J. Robbins *et al.*, The variability of crater identification among expert and community crater analysts. *Icarus* **234**, 109-131 (2014). doi: 10.1016/j.icarus.2014.02.022
64. K. A. Holsapple, The scaling of impact processes in planetary sciences. *Annu. Rev. Earth Planet Sci.* **21**, 333-373 (1993).
65. K. R. Housen, K. A. Holsapple, Ejecta from impact craters. *Icarus* **211**, 856-875 (2011). doi: 10.1016/j.icarus.2010.09.017
66. W. B. McKinnon, P. M. Schenk, Estimates of comet fragment masses from impact crater chains on Callisto and Ganymede. *Geophys. Res. Lett.* **22**, 1829-1832 (1995).
67. K. N. Singer, S. A. Stern, On the provenance of Pluto's nitrogen ($N_2$). *The Astrophysical Journal Letters* **808**, L50 (2015). doi: doi:10.1088/2041-8205/808/2/L50
68. W. C. Fraser *et al.*, The Kuiper belt luminosity function from m=21 to 26. *Icarus* **195**, 827-843 (2008). doi: 10.1016/j.icarus.2008.01.014
69. D. R. Davis, D. D. Durda, F. Marzari, A. Campo Bagatin, R. Gil-Hutton, in *Asteroids III*, W. F. Bottke, A. Cellino, P. Paolicchi, R. P. Binzel, Eds. (University of Arizona Press, Tucson, 2002), pp. 545–558.

70. S. A. Stern, J. E. Colwell, Collisional erosion in the primordial Edgeworth-Kuiper belt and the generation of the 30-50 au Kuiper gap. *Astrophys J.* **490**, 879-882 (1997).
71. D. R. Davis, P. Farinella, Collisional evolution of Edgeworth-Kuiper belt objects. *Icarus* **125**, 50-60 (1997). doi: 10.1006/icar.1996.5595
72. S. J. Kenyon, J. X. Luu, Accretion in the early Kuiper belt. II. Fragmentation. *Astron. J.* **118**, 1101-1119 (1999). doi: 10.1086/300969
73. R. Gomes, H. F. Levison, K. Tsiganis, A. Morbidelli, Origin of the cataclysmic late heavy bombardment period of the terrestrial planets. *Nature* **435**, 466-469 (2005). doi: 10.1038/nature03676
74. H. J. Melosh, *Planetary surface processes*. (Cambridge University Press, Cambridge, 2011).
75. A. J. Dombard, W. B. McKinnon, Elastoviscoplastic relaxation of impact crater topography with application to Ganymede and Callisto. *J. Geophys. Res.* **111**, E01001 (2006). doi: 10.1029/2005JE002445
76. M. T. Bland, K. N. Singer, W. B. McKinnon, P. M. Schenk, Viscous relaxation of Ganymede's impact craters: Constraints on heat flux. *Icarus* **296**, 275-288 (2017). doi: 10.1016/j.icarus.2017.06.012
77. K. N. Singer, M. T. Bland, P. M. Schenk, W. B. McKinnon, Relaxed impact craters on Ganymede: Regional variation and high heat flows. *Icarus* **306**, 214-224 (2017). doi: 10.1016/j.icarus.2018.01.012
78. G. C. Collins *et al.*, in *Planetary tectonics*, T. R. Watters, R. A. Schultz, Eds. (Cambridge University Press, Cambridge, 2010), pp. 264-350.
79. R. A. Beyer *et al.*, Charon tectonics. *Icarus* **287**, 161-174 (2017). doi: 10.1016/j.icarus.2016.12.018
80. R. A. Beyer *et al.*, paper presented at the 48th Annual American Astronomical Society Division for Planetary Sciences Meeting, Pasadena, CA, 1 October.
81. A. D. Howard *et al.*, Present and past glaciation on Pluto. *Icarus* **287**, 287-300 (2017). doi: 10.1016/j.icarus.2016.07.006
82. W. M. Grundy *et al.*, Surface compositions across Pluto and Charon. *Science* **351**, (2016). doi: 10.1126/science.aad9189
83. A. M. Earle, R. P. Binzel, Pluto's insolation history: Latitudinal variations and effects on atmospheric pressure. *Icarus* **250**, 405-412 (2015).
84. K. N. Singer *et al.*, paper presented at the 47th Annual Lunar and Planetary Science Conference, Houston, TX, 1 March 2016, #2276.
85. W. B. McKinnon *et al.*, paper presented at the 48th Annual Lunar and Planetary Science Conference, Houston, TX, 1 March 2017, #2854.
86. S. J. Robbins *et al.*, Investigation of Charon's craters with abrupt terminus ejecta, comparisons with other icy bodies, and formation implications. *Journal of Geophysical Research: Planets* **123**, 20-36 (2018). doi: 10.1002/2017je005287
87. M. T. Bannister *et al.*, OSSOS. IV. Discovery of a dwarf planet candidate in the 9:2 resonance with Neptune. *Astron. J.* **152**, (2016). doi: 10.3847/0004-6256/152/6/212
88. K. Volk *et al.*, OSSOS III—resonant trans-neptunian populations: Constraints from the first quarter of the Outer Solar System Origins Survey. *Astron. J.* **152**, (2016). doi: 10.3847/0004-6256/152/1/23
89. C.-Y. Liu *et al.*, Search for sub-kilometre trans-neptunian objects using CoRoT asteroseismology data. *Mon. Not. R. Astron. Soc.* **446**, 932-940 (2015).

90. C.-Y. Liu, A. Doressoundiram, F. Roques, H.-K. Chang, S. Chaintreuil, paper presented at the Euro. Planet. Sci. Cong., Riga, Latvia, 1 September, #EPSC2017-2237.
91. Z.-W. Zhang *et al.*, The TAOS project: Results from seven years of survey data. *Astron. J.* **146**, (2013).
92. M. Jutzi, E. Asphaug, The shape and structure of cometary nuclei as a result of low-velocity accretion. *Science* **348**, 1355-1358 (2015). doi: 10.1126/science.aaa4747
93. M. J. Mumma, P. R. Weissman, S. A. Stern, J. I. Lunine, in *Protostars and planets III,* E. H. Levy, Ed. (University of Arizona Press, Tucson, 1993), pp. 1177-1252.
94. H. Rickman *et al.*, Comet 67P/Churyumov-Gerasimenko: Constraints on its origin from OSIRIS observations. *A & A* **583**, (2015). doi: 10.1051/0004-6361/201526093
95. O. L. White *et al.*, Geological mapping of Sputnik Planitia on Pluto. *Icarus* **287**, 261-286 (2017). doi: 10.1016/j.icarus.2017.01.011
96. LROC WAC global morphologic map. **http://wms.lroc.asu.edu/lroc/view_rdr/WAC_GLOBAL**, (2011).
97. W. B. McKinnon, K. N. Singer, paper presented at the 42nd Annual American Astronomical Society Division for Planetary Sciences Meeting, Pasadena, CA, 1 October 2010, #984.
98. R. Strom *et al.*, The inner solar system cratering record and the evolution of impactor populations. *Reasearch in Astronomy and Astrophysics* **15**, 407-434 (2015). doi: 10.1088/1674–4527/15/3/009
99. S. Marchi, W. F. Bottke, D. A. Kring, A. Morbidelli, The onset of the lunar cataclysm as recorded in its ancient crater populations. *Earth Planet. Sci. Lett.* **325**, 27-38 (2012). doi: 10.1016/j.epsl.2012.01.021
100. N. Schmedemann *et al.*, The cratering record, chronology and surface ages of (4) Vesta in comparison to smaller asteroids and the ages of HED meteorites. *Planet. Space Sci.* **103**, 104-130 (2014).
101. I. B. Aban, M. M. Meerschaert, A. K. Panorska, Parameter estimation for the truncated pareto distribution. *Journal of the American Statistical Association* **101**, 270-277 (2006).
102. S. J. Robbins *et al.*, Revised recommended methods for analyzing crater size-frequency distributions. *Meteoritics and Planetary Science* **53**, 891-931 (2018). doi: 10.1111/maps.12990
103. S. Marchi *et al.*, Small crater populations on Vesta. *Planet. Space Sci.* **103**, 96-103 (2014). doi: 10.1016/j.pss.2013.05.005
104. D. A. Minton, R. Malhotra, Dynamical erosion of the asteroid belt and implications for large impacts in the inner Solar System. *Icarus* **207**, 744-757 (2010).
105. D. W. G. Arthur, A. P. Agnieray, R. A. Horvath, C. A. Wood, C. R. Chapman, The system of lunar craters, Quadrant I. *Communications of the Lunar and Planetary Laboratory* **2**, 71 (1964).